\documentclass{article}
\usepackage{amsmath}
\usepackage{amsfonts}
\usepackage{cite}
\usepackage{hyperref}
\numberwithin{equation}{section}

\def\p{\partial}
\def\O{\mathcal{O}}
\def\L{\mathcal{L}}
\def\a{\alpha}
\def\b{\beta}
\def\g{\gamma}
\def\r{\rightarrow}

\def\H{\mathcal{H}}
\def\P{\mathcal{P}}
\def\l{\lambda}
\def\d{\delta}
\def\s{\sigma}
\def\e{\epsilon}

\def\k{\kappa}
\def\J{\mathcal{J}}
\def\H{\mathcal{H}}
\def\V{\mathcal{V}}
\def\ts{\tilde{\sigma}}
\def\X{\mathcal{X}}
\def\etao{\eta_{\mbox{\tiny{$\mathcal{O}$}}}}
\def\baretao{\bar{\eta}_{\mbox{\tiny{$\mathcal{O}$}}}}
\def\medpi{\mbox{\small{$\varPi$}}}

\newcommand{\be}{\begin{equation}}
\newcommand{\ee}{\end{equation}}
\newcommand{\bea}{\begin{eqnarray}}
\newcommand{\eea}{\end{eqnarray}}
\newcommand{\bi}{\begin{itemize}}
\newcommand{\ei}{\end{itemize}}

\title{A definition of primary operators in $J\bar T$ - deformed CFTs\vspace{7mm}}

\begin{document}

\author{
Monica Guica \vspace{0mm} \\
\\\vspace{1mm}
\emph{\small Universit\'e Paris-Saclay, CNRS, CEA,} 
\emph{\small Institut de Physique Th\'eorique, 91191 Gif-sur-Yvette, France}}

\date{}
\maketitle

\begin{abstract}

\vskip8mm

\noindent $J\bar T$ - deformed CFTs provide an interesting example of non-local, yet UV-complete two-dimensional QFTs that are entirely solvable. They have been recently shown to possess an infinite set of symmetries, which are a continuous deformation of the Virasoro-Kac-Moody symmetries of the seed CFT.  In this article, we put forth a definition of primary operators in $J\bar T$ - deformed CFTs on a cylinder, which are singled out by  having CFT-like momentum-space commutation relations with the symmetry generators in the decompatification limit. We show - based on results we first derive for the case of $J^1 \wedge J^2$ - deformed CFTs - that  all correlation functions of such  operators in the $J\bar T$ - deformed CFT can be computed exactly in terms of the  correlation functions of the undeformed CFT and are crossing symmetric in the plane limit. In particular, two and three-point functions are simply given by the corresponding momentum-space correlator in the undeformed CFT,  with all dimensions replaced by particular momentum-dependent conformal dimensions. Interestingly, scattering amplitudes off the near-horizon of extremal black holes are known to take a strikingly similar form.   

\end{abstract}

\tableofcontents

\section{Introduction}

A long-standing challenge in  holography has been to find the microscopic description of \emph{generic} extremal and non-extremal black holes,  whose near-horizon region does not contain an AdS$_3$ factor. One can obtain important clues into the nature of the holographic dual by studying the asymptotic symmetries of the near-horizon backgrounds of interest \cite{Brown:1986nw}, as well as scattering amplitudes \cite{
Maldacena:1997ih,Maldacena:1996ix}, which in principle give access to the symmetries \cite{Strominger:1997eq} and, respectively, the correlation functions of the dual theory. 

To date, the most progress has been made in the case of extremal black holes, whose near-horizon region universally contains \cite{Kunduri:2007vf} a factor known as warped AdS$_3$: a deformation of AdS$_3$ that preserves $SL(2,\mathbb{R}) \times U(1)$ isometry. The asymptotic symmetries of this space-time enhance the $U(1)$ factor to one copy of the Virasoro algebra, leading to the so-called Kerr/CFT proposal \cite{Guica:2008mu}, which states that the near-horizon dynamics of extremal black holes is described by a chiral half  of a two-dimensional CFT, where the chirality is due to taking the strict extremal limit \cite{Balasubramanian:2009bg}. 

The study of scattering off the near-extremal black hole geometry naively  appears to corroborate this claim,  since it leads to a (scalar) momentum-space two-point function of the form \cite{Bredberg:2009pv} 

\be
\mathcal{G}_{\mbox{\tiny{$T_{L,R}$}}}(p,\bar p) \sim T_L^{2h(\bar p)-1}  e^{- \frac{p}{2T_L}} \left|\Gamma\left(h(\bar p) + \frac{i p}{2\pi T_L}\right)\right|^2 \; \times \;  T_R^{2h (\bar p)-1}  e^{- \frac{\bar p}{2T_R}} \left|\Gamma\left(h(\bar p) + \frac{i \bar p}{2\pi T_R}\right)\right|^2 \label{bh2pf}
\ee
which precisely corresponds to the momentum-space two-point function in a two-dimensional CFT at left and, respectively, right-moving temperatures $T_{L,R}$. Similar results hold for higher-point functions \cite{Becker:2010jj}. 

This conclusion may nevertheless be  a bit too fast, because the operator dimensions that appear in the above formula, which are extracted from  the solution to the scalar wave equation in the near-horizon geometry, depend  explicitly on the momentum, $\bar p$, along the  $U(1)$ direction. This fact immediately implies that the ``CFT'' in the Kerr/CFT correspondence cannot be a standard, local CFT\footnote{
 Consequently, the asymptotic symmetry results of \cite{Guica:2008mu} (or, more rigorously, \cite{Compere:2014bia}) should not be interpreted as suggesting that the dual theory is a standard CFT, but rather that non-local theories can posses Virasoro symmetry, a fact that was recently proven in \cite{Guica:2021pzy,Guica:2020eab}. }. Indeed, a detailed study of the near-horizon geometry \cite{El-Showk:2011euy} reveals that the holographic dual is instead a two-dimensional analogue of a dipole theory \cite{Bergman:2000cw},  obtained by deforming the CFT by a finely-tuned set of irrelevant  operators 
that preserve the left $SL(2,\mathbb{R})$ symmetry and lead  to a UV-complete theory. This structure of the irrelevant deformation implies that the resulting theory - sometimes called a ``dipole CFT'' - is local and conformal on the left, but non-local on the right. Operators in these theories are best described in a mixed position - momentum basis, and the left-moving piece of their correlation functions has a form dictated by the $SL(2,\mathbb{R})_L$ conformal symmetry; however, the left conformal dimensions  generically  depend on the right-moving momentum \cite{Guica:2010sw}.

The fact that the dual theory is non-local raises some immediate questions regarding \eqref{bh2pf}. While the left-moving piece of this correlator is meaningful -  since it corresponds to the Fourier transform of a local, conformal two-point function (albeit with a dimension that   depends on the momentum along the other direction) - its right-moving piece, which resides entirely on the non-local direction, appears to have no intrinsic meaning, since it can always be rescaled by an arbitrary function of $\bar p$ \cite{vanRees:2012cw}.  
%
 The question that we would like to address in this article is: to what extent is an expression of the form  \eqref{bh2pf} in such a non-local theory meaningful?

That this correlator may actually be meaningful is suggested by the example of dipole-deformed field-theories, whose  non-locality is extremely constrained by a star product structure. 
 Operators in this theory and the  allowed counterterms  must respect this structure, which effectively  makes the theory renormalizable \cite{Dasgupta:2001zu}, if the original QFT was so. Thus, in presence of  additional structures that constrain the non-locality, one may hope to be able to assign an unambiguous meaning to an expression such as \eqref{bh2pf}. 

The main goal of this article is to identify  such additional structures for the case of two-dimensional non-local QFTs that may model the microscopic physics of generic extremal black holes. Specifically, we will be working with the example of $J\bar T$ - deformed CFTs \cite{Guica:2017lia} -  a class of \emph{universal} irrelevant deformations of two-dimensional  CFTs by an operator that is bilinear in the stress tensor and a $U(1)$ current, which lead to a UV-complete QFT. These theories belong to the more general class of  Smirnov-Zamolodchikov irrelevant current-current deformations \cite{Smirnov:2016lqw}, of which the $T\bar T$ deformation \cite{Smirnov:2016lqw,Cavaglia:2016oda}  is a particularly rich and interesting example \cite{
Dubovsky:2012wk,Dubovsky:2013ira,Dubovsky:2018bmo,Giveon:2017nie}. While $J\bar T$ - deformed CFTs are not exactly a model for the Kerr/CFT correspondence because the deformation is double-trace - and thus corresponds to AdS$_3$ with mixed boundary conditions \cite{Bzowski:2018pcy} - a single-trace variant \cite{Apolo:2018qpq,Chakraborty:2018vja} of this deformation is. Notwithstanding, $J\bar T$ - deformed CFTs do appear to posses the  correct non-local QFT structures 
 that are relevant for the Kerr/CFT correspondence,  
 in addition to being highly tractable. The concrete question that we would like to address in this article  is  whether a formula such as \eqref{bh2pf} can be made sense of in the context of $J\bar T$ - deformed CFTs. 

This question has two aspects. The first is that one should find a basis of operators  for which the correlation functions are expected to take a simple form. In usual CFTs, these are local operators -  in fact, primary operators,  if we want  the higher-point functions to also be nicely constrained. In $J\bar T$ - deformed CFTs, the primary constraint is  easily imposed on the local  left-moving side and yields correlation functions that are consistent with $SL(2,\mathbb{R})$ invariance,  the only change being that the left conformal dimensions and  $U(1)$ charges are shifted from their undeformed values  $\tilde h, \tilde q$ by  a momentum-dependent contribution  \cite{Guica:2019vnb}

\be
h(\bar p) = \tilde h + \l \tilde q \bar p + \frac{\l^2 k}{4} \bar p^2 \;, \;\;\;\;\; q(\bar p ) = \tilde q + \frac{\l k}{2} \bar p \label{momdepdimsjtb}
\ee
where  $\l$ is the deformation parameter. 
Our (more non-trivial) goal  is to find a similar constraint that  fixes the right-moving dependence of the correlators. 

 The other aspect concerns the method used to compute the correlation functions. Given the definition of the deformation in terms of an irrelevant flow, conformal perturbation theory comes in naturally, and this was the method used in  \cite{Guica:2019vnb} to analyse $J\bar T$ correlators, and in \cite{Cardy:2019qao} (see also \cite{
Aharony:2018vux,Rosenhaus:2019utc,He:2019vzf}) for the more involved $T\bar T$ case. However, note that the correlation functions computed with this method are UV divergent, and so need to be regulated and renormalized; however,  it is \emph{a priori} not clear whether any choice of UV regulator and which choices of counterterms would be allowed. 
 


In this article, we approach  the computation of $J\bar T$ correlators in a different way, which circumvents the issue of UV divergences and understanding what are the allowed counterterms. Instead, we rely on the flow equation with respect to $\l$ satisfied by the energy eigenstates, the symmetry generators and an appropriately-defined set of operators  to construct the correlation functions of interest. Our approach is similar in spirit to the one used in \cite{Kruthoff:2020hsi} to discuss $T\bar T$ correlators, though our basis of operators is different and the correlation functions we compute are manifestly finite throughout, as well as fully explicit. 

 More concretely, to fix the right-moving part of the correlator, we use the fact that  $J\bar T$ - deformed CFTs were recently shown to possess an infinite-dimensional ``pseudo-conformal'' symmetry \cite{infconf}, implemented - at the classical level - by field-dependent generalizations of conformal and affine $U(1)$ transformations 

\be
v \;  \r \; \bar f(v)   \;, \;\;\;\;\;\;\;\; v\equiv V-\l \phi\label{fdeptr}
\ee
where $V$ is the right-moving coordinate and  $\phi$ is the bosonisation of the $U(1)$ current (with its zero mode removed, if working on the cylinder).  The corresponding symmetry generators have been constructed  in \cite{infconf,Guica:2020eab} at the classical level, and in \cite{Guica:2021pzy} at the full quantum level. The quantum construction relied on the existence of an alternate basis for these symmetries, denoted as the ``flowed'' representation, in which the generators  - denoted with a ~$\widetilde{}$~ -   are simply defined to flow in the same way as the energy eigenstates. By construction, they satisfy a Virasoro-Kac-Moody algebra \cite{LeFloch:2019rut}; more non-trivially, they can also be shown to be conserved.  A similar construction holds on the left-moving side. The generators of the left conformal and right pseudo-conformal \eqref{fdeptr} symmetries are  given in terms of the flowed symmetry generators by

%

\be
L_n = \tilde L_n + \l H_R \tilde J_n + \frac{\l^2 k H_R^2}{4}\, \d_{n,0}\;, \;\;\;\;\;\;\bar L_n = \tilde{\bar L}_n + \l : H_R \tilde{\bar{J}}_n : + \frac{\l^2 k H_R^2}{4}\, \d_{n,0} \nonumber
\ee

\be
J_n = \tilde J_n + \frac{\l k}{2} H_R \, \d_{n,0} \;, \;\;\;\;\;\;\; \bar J_n = \tilde{\bar J}_n + \frac{\l k}{2} H_R \, \d_{n,0} 
\ee
where  the $:\;:$ denote normal ordering, i.e. $\tilde{\bar{J}}_n$ is to the right of $H_R$ for $n>0$, and to the left for $n<0$.
Note the above relation resembles a `spectral flow' by the right-moving Hamiltonian, $H_R$.
The algebra of these generators is Virasoro-Kac-Moody on the left and a non-linear  deformation of it on the right; also, the left and right generators do not commute.

It is interesting to ask whether this deformed (Virasoro-Kac-Moody)${}^2$ symmetry  can help us fix the form of correlation functions, as it does in usual CFTs.

In two-dimensional CFTs, one can define primary operators  - of dimension $(h, \bar h)$  - either through their transformation properties 
 under finite conformal transformations $z \r z'(z), \bar z \r \bar z'(\bar z)$

\be
\O' (z',\bar z') = (\p_z z')^{-h} (\p_{\bar z} \bar z')^{-\bar h} \O(z,\bar z) \label{primcft}
\ee 
or through their commutation relations with the Virasoro generators
%

\be
[L_n, \O(z)] = h(n+1) z^n \O + z^{n+1} \p_z \O \;, \;\;\;\;\;\; n \geq -1 \label{wardidz}
\ee
and similarly on the right. For $n= \pm 1, 0$ this relation, together with the  $SL(2,\mathbb{R})$  invariance of the vacuum,  completely fixes the form of two and three-point functions, and highly constrains higher-point ones.

In $J\bar T$ - deformed CFTs, it is not clear how to define an analogue of \eqref{primcft}. One may naively attempt to simply replace $\bar z$ in \eqref{primcft} by its field-dependent counterpart $\bar z - \l \phi$; however, this is in tension with the known fact that the left-moving dimensions  depend on the right-moving momentum as  in  \eqref{momdepdimsjtb}, which forces one to treat at least the right-movers  in momentum space. As we will show, it is nevertheless possible to come rather close to a definition of `primary' operators in $J\bar T$ - deformed CFTs that is  analogous to the  momentum-space counterpart of \eqref{wardidz}.   The correlation functions of the resulting operators
%
 are  completely fixed 
 in terms of the corresponding undeformed CFT correlators and, interestingly, take a form that  highly resembles \eqref{bh2pf}. To what extent our proposal is `the' correct definition of primary operators in these non-local CFTs is  left for future investigations.  
 
This article is organised as follows. In section \ref{jjbwarm}, we study a simple toy model for the $J\bar T$ deformation
, which also exhibits two sets of symmetry generators related by an operator-dependent spectral flow. For this example we show, using only flow equations and the interplay of  the symmetry generators,  that general deformed primary correlators   are completely determined by their undeformed counterparts. 
In section \ref{primjtb}, we apply the same technique to compute general correlation functions in $J\bar T$ - deformed  CFTs, making  appropriate adjustments for the non-locality of the model. We conclude with a discussion in section \ref{disc}. Various technical details of the symmetry algebras are detailed in the appendices, as well as a very explicit realisation 
of the toy model of section \ref{jjbwarm} in terms of deformed free bosons.

\section{$J^1\wedge J^2$ warmup \label{jjbwarm}}

In this section, we would like to ``warm up'' for the construction of primary operators in $J\bar T$ - deformed CFTs by studying a simple toy model that also exhibits two possible bases for the symmetry generators, related by an operator-dependent spectral flow. This model is the so-called $J^1 \wedge J^2$ deformation of a two-dimensional CFT, which is the simplest possible Smirnov-Zamolodchikov  deformation - built from two $U(1)$ currents - and also corresponds to a two-dimensional toy model for a four-dimensional gauge theory in presence of a $\theta$ term \cite{Cardy:1981qy}.

 Since the  $J^1 \wedge J^2$ deformation is exactly marginal,  all the powerful tools of conformal symmetry are still  applicable to the deformed theory; in particular, it is very clear what the primary operators are. Throughout this section, we will  make an effort to treat this deformation in a language that is also applicable to the 
 $J\bar T$  deformation, which will provide a very useful guiding principle for how to proceed  when the conformal symmetry is deformed to the non-local, field-dependent symmetries \eqref{fdeptr}.

\subsection{Brief review of the  $J^1 \wedge J^2$  deformation \label{brev}}

We consider a one-parameter $(\l)$ family of two-dimensional CFTs  that posses  two global $U(1)$ conserved currents,  $J^{1,2}$. The actions describing the various members of this family are related via the flow equation


\be
\frac{\p S}{\p \l} = -  \int d^2 x\, e^{\a\b} \left( J^1_\a J^2_\b \right)_\l 
\ee
where the bilinear operator appearing on the right-hand side is defined via point-splitting \cite{Smirnov:2016lqw} and  the current components are  computed in the deformed CFT. The deformed spectrum has been understood in \cite{Cardy:1981qy,bulch}, and certain aspects of correlation functions have been analysed in \cite{Cardy:2019qao}. In this subsection, we will review some of these results, in a language that parallels the $J\bar T$ analysis of \cite{Guica:2021pzy}. 

\subsubsection*{Classical analysis}

 It is useful to first understand the effect of the $J^1 \wedge J^2$ deformation at the classical level. This is perhaps simplest to present in   Hamiltonian language. We thus consider a Hamiltonian density $\H(\pi_i, \phi_i)$, which admits at least two $U(1)$ symmetries that we associate to shifts in two scalars, $\phi_{1,2}$. As a result, the Hamiltonian  depends on these two canonical variables only through their spatial derivatives, $\phi_{1,2}'$. The two  shift currents have components 
\be
J^a_t = \pi^a \, \sqrt{k}\;, \;\;\;\;\;\; J^a_\s = \p_{\phi'_a} \H \, \sqrt{k}\;, \;\;\;\;\; a=1,2 \label{shiftc}
\ee
where we have allowed for an arbitrary level\footnote{Even if $k$ is an  anomaly coeficient, for bosons  it  appears already at the level of the classical Poisson brackets. This allows one to understand many properties of the deformation with just a classical analysis. }, $k$. We will also consider the topologically conserved currents 

\be
\tilde J^a_t = \phi'^a \,  \sqrt{k} \;, \;\;\;\;\;\;\; \tilde J^a_\s = \p_{\pi_a} \H\, \sqrt{k} \label{topc}
\ee
and will assume that in the undeformed CFT, with Hamiltonian density $\H^{(0)} (\pi_i, \phi_i)$, the combinations $J^a_\a \pm   \tilde J^a_\a$ are (anti)chiral, which implies that 

\be
\p_{\pi^a} \H^{(0)} =  \pi^a \;, \;\;\;\;\;\; \p_{\phi'_a} \H^{(0)} = \phi_a' \label{chirh0}
\ee
The flow equation obeyed by the deformed Hamiltonian  reads (in the convention $\e_{t\s} = \e^{\s t} = 1$)

\be
\p_\l \H = \e^{\a\b} J^1_\a J^2_\b =  k( \pi_2 \p_{\phi_1'} \H- \pi_1 \p_{\phi_2'} \H) \label{hamfljjb}
\ee
This equation can be solved  by making the Ansatz 

\be
\H(\l) = \widetilde \H (\l) + \frac{\l^2 k^2}{2} (\pi_1^2 + \pi_2^2) + \l k (\phi_1'\pi_2 -\phi_2' \pi_1   ) \label{defhamjjb}
\ee
which implies that $\widetilde \H(\l)$ satisfies the equation

\be
\p_\l \tilde \H  =   k \pi_2 (\p_{\phi_1'} \widetilde \H - \phi_1')- k \pi_1 (\p_{\phi_2'}\widetilde \H - \phi_2')
\ee
This is solved by  $\widetilde \H(\l) = \H^{(0)}$, using the initial condition \eqref{chirh0}.  One can easily check, following e.g. \cite{Jorjadze:2020ili},  that the  stress tensor computed from the  resulting deformed  Hamiltonian \eqref{defhamjjb} is both  symmetric and traceless, and thus the deformed theory remains a CFT. In this theory, it is useful we introduce the left/right Hamiltonian currents ($\P$ is the momentum density)
 
 \be
 \H_{L,R} \equiv \frac{\H\pm \P}{2} \;, \;\;\;\;\;\;\; \P = \sum_i \pi_i \phi'_i
 \ee
We can now use the deformed Hamiltonian \eqref{defhamjjb} 
 and the definitions \eqref{shiftc}, \eqref{topc} to compute the components of the deformed conserved currents.  A basis for the currents that are now (anti)chiral is given by 

\be
\J^1_{L,R} = \frac{\sqrt{k}}{2} (\pi_1 \pm \phi'_1 \pm \l k \pi_2)\;, \;\;\;\;\;\;\; \J^2_{L,R} = \frac{\sqrt{k}}{2} (\pi_2 \pm \phi'_2 \mp  \l k \pi_1) \label{chircurr}
\ee
where the above expressions represent their time components and, by definition, $\J^a_{L,\s} = \J_{L,t}^a$, $\J^a_{R,\s} = - \J^a_{R,t}$.  The  Poisson brackets of the currents in this basis  are diagonal and $\l$ - independent 

\be
\left\{ \J^a_{L} (\s), \J^b_L (\tilde \s) \right\} = -\left\{ \J^a_{R} (\s), \J^b_R (\tilde \s) \right\} = \frac{k}{2} \d'(\s-\tilde \s) \;, \;\;\;\;\;\; \left\{ \J^a_{L} (\s), \J^b_R (\tilde \s) \right\} = 0
\ee
and their commutators with the deformed Hamiltonian currents $\H_{L,R} (\l)$ yield the standard Witt-Kac-Moody algebra between (anti)chiral currents in a CFT. 
%
It is also interesting to note that the combinations

\be
\hat \H_L \equiv \H_L - \frac{1}{k} \sum_a (\J_L^a)^2  \;, \;\;\;\;\;\;\;\hat \H_R \equiv \H_R -  \frac{1}{k} \sum_a (\J_R^a)^2  
\ee
are independent of $\l$. Thus, they equal the corresponding quantities in the undeformed CFT, which are nothing but  the spectral-flow-invariant piece of the Hamiltonian currents. This structure is reminiscent of that of the left-moving  Hamiltonian in $J\bar T$ - deformed CFTs \cite{Guica:2021pzy}. 

One final observation that  will be useful shortly is that, if we define the total currents  $\J^a_\a \equiv \J^a_{L,\a} +\J^a_{R,\a}$, then  the deforming operator can also  be written  in terms of them as

\be
\O_{J^1\wedge J^2} = \e^{\a\b} J^1_\a J^2_\b = 2 (\J_L^1 \J^2_R - \J^1_R \J^2_L ) =  \e^{\a\b} \J^1_\a \J^2_\b \label{flopid}
\ee

\subsubsection*{Quantum analysis}

We now move on to the quantum theory, and place the $J^1 \wedge J^2$ - deformed CFT  on a cylinder of radius $R$. Following \cite{Smirnov:2016lqw}, we  consider eigenstates $|n_\l\rangle$ of the  energy and the charge, whose shift and, respectively, winding  charges are 
\be
n^a = \int d\s \, \langle J^a_t \rangle \;, \;\;\;\;\;\;\;\; w_a = \int d\s \, \langle \tilde J^a_t \rangle
\ee
Since the deformation is integrable, it does not change the Hilbert space of states on the cylinder, but it induces a flow of the energy eigenstates, of the form

\be
\p_\l |n_\l\rangle = \mathcal{X}_{\mbox{\tiny{$J \bar J$}}} |n_\l\rangle \label{jjbflee}
\ee
where $\X_{\mbox{\tiny{$J \bar J$}}}=-\X^\dag_{\mbox{\tiny{$J \bar J$}}}$ is a well-defined  operator acting on the Hilbert space, which we  will determine shortly.

We would now like to understand how the energies and \emph{chiral} charges $q^a, \bar q^a$ - i.e., the charges associated to the zero modes
\be
\mathcal{Q}^a = \int d\s \, \J_L^a \;, \;\;\;\;\;\; \bar{\mathcal{Q}}^a= \int d\s \, \J_R^a \label{chirch}
\ee
of the currents \eqref{chircurr} - depend on  $\l$. Using first order quantum-mechanical perturbation theory and  the factorization properties of the Smirnov-Zamolodchikov operator in energy eigenstates, we find
\bea
\p_\l E_n^\l &=& \langle n_\l | \p_\l H |n_\l\rangle = R \left(  \langle n_\l| J^1_\s |n_\l \rangle \langle n_\l |J^2_t |n_\l \rangle - \langle n_\l| J^1_t | n_\l \rangle  \langle n_\l| J^2_\s |n_\l \rangle \right) \nonumber \\
&  & \hspace{0.5cm} = \; \frac{1}{R} \left[ n_2 (w_1 + \l k n_2) -  n_1 (w_2-\l k n_1) \right]
\eea
where $\p_\l H$ is the spatial integral of \eqref{hamfljjb} over the circle and we used \eqref{shiftc} to compute the spatial components $J^a_\s$.  Since the shift and winding charges defined above are quantized, and thus cannot flow with $\l$, this equation immediately integrates to the following expression for the deformed energies 

\be
E_n^\l = E_n^{(0)} + \frac{\l^2 k}{2 R} (n_1^2+n_2^2) + \frac{\l }{R} (w_1 n_2 - w_2 n_1)
\ee
While this expression is entirely analogous to \eqref{defhamjjb},  note that it does not immediately follow from it. 

The charges associated to the chiral conserved  currents \eqref{chircurr}  are given by 

\be
q^a = \tilde q^a + \frac{\l k}{2} \e^{ab} n_b \;, \;\;\;\;\;\;  \bar q^a = \tilde{\bar q}^a - \frac{\l k}{2} \e^{ab} n_b \label{defchjjb}
\ee
where $\tilde q_a , \tilde{\bar q}_a \equiv (n_a \pm w_a)/2 $ stand for the undeformed (anti)chiral charges  and $\e_{12} = \e^{12} =1$.
Using the state-operator correspondence to map the energies of eigenstates  on the cylinder to the conformal dimensions of local operators on the plane, we  find that  the spectrum of left/right conformal dimensions in the $J^1\wedge J^2$ - deformed CFT depends on $\l$ as
 
\be
h = \tilde h + \l \e_{ab} \tilde q^a n^b + \frac{\l^2 k }{4}\, n_a n^a \;, \;\;\;\;\;\; \bar h = \tilde{\bar h} - \l \e_{ab} \tilde{\bar{q}}^a n^b + \frac{\l^2 k}{4}\, n_a n^a \label{defdimsjjb}
\ee
where $\tilde h$, $\tilde {\bar h} = (E_n^{(0)} \pm P_n) R$ are the undeformed conformal dimensions. 

Thus, the effect of the $J^1 \wedge J^2$ deformation on the spectrum of energies or, equivalently, on the local operator dimensions is precisely that of a simultaneous spectral flow  in the two $U(1)$ directions, with charge-dependent parameters $ \eta^a = - \bar \eta^a =  \l \e^{ab} n_b$ that are opposite on the left and the right. Note that  the total momentum charge $n^a = q^a + \bar q^a$ is unaffected. One can  easily check that $\hat h = h - (q_a q^a)/k$ is left invariant, as expected.

\subsection{Flow of the states and of the symmetry generators}

We would now like to determine  the form of the operator $\X_{\mbox{\tiny{$J \bar J$}}}$ entering the flow equation   \eqref{jjbflee} for the $J^a \wedge J^b$ - deformed energy eigenstates at least in the classical limit, in analogy with the results of  \cite{Guica:2020eab} for $J\bar T$.  Using first order quantum-mechanical perturbation theory and assuming the CFT degeneracies are dealt with,  this operator can be read off from

\be
\p_\l |n\rangle_\l = \sum_{m \neq n} \frac{\langle m_\l| \p_\l H|n_\l\rangle}{E_n^\l-E_m^\l} \,  |m_\l\rangle \label{foqmpt}
\ee
where $\p_\l H$ is the spatial integral of \eqref{hamfljjb}, performed on the $t=0$ slice\footnote{As explained in \cite{Guica:2020eab}, for $t \neq 0$ the flow operator will receive additional contributions proportional to $\p_\l E_n^\l$. }

\be
\p_\l H = \int d\s  \, \e^{\a\b} J^1_\a J^2_\b = \int d\s \, \e^{\a\b} \J^1_\a \J^2_\b 
\ee
where we used \eqref{flopid}. To obtain  a useful expression for $\X_{\mbox{\tiny{$J \bar J$}}}$, we  follow  the steps outlined in \cite{Kruthoff:2020hsi} for the case of $T\bar T$. 
This involves splitting the two  current insertions in the deforming operator using a $\d$ function, which is subsequently rewritten in terms of the Green's function  on the cylinder, which satisfies 
\be
\p_\s G(\s-\ts) = \d (\s-\ts) - \frac{1}{R}
\ee
After these manipulations, we obtain 

\be
\p_\l H = \frac{2}{R} \, \e_{ab}\mathcal{Q}^a \bar{\mathcal{Q}}^b - \p_t \int d\s d\tilde \s \, G(\s-\tilde \s) \J^1_t (\s) \J^2_t (\tilde \s)
\ee
where $\mathcal{Q}_a, \bar{\mathcal{Q}}_a$ are the  chiral charge operators  \eqref{chirch}.  The first term can also be written as a total time derivative by bosonising the currents  $\J^a_{L,R} = \p_\pm  \varphi_{L,R}^a $  and noting that the zero modes  of the chiral scalars thus introduced  satisfy

\be
[  H, \varphi_{L,0}^a ] =- i \mathcal{Q}^a  \;, \;\;\;\;\;\; [  H, \varphi_{R,0}^a ] =- i \bar{\mathcal{Q}}^a
\ee
Then,  $2 \e_{ab}\mathcal{Q}^a \bar{\mathcal{Q}}^b= [H, i \e_{ab}( \varphi_{L,0}^a \bar{\mathcal{Q}}^b + \mathcal{Q}^a \varphi_{R,0}^b) ] = -i \frac{d}{dt} (\ldots)$. The next step is to use an integral representation of the denominator in \eqref{foqmpt} to rewrite it as 

\be
\p_\l |n_\l\rangle = -i \sum_{m\neq n} \int_{-\infty}^0 dt \, e^{t \e } |m_\l\rangle \langle m_\l | \p_\l H (t) |n_\l\rangle  
\ee 
where $\e>0$ is an infinitesimal regulator. Performing the integral and taking $\e \r 0$, we find

\be
\p_\l |n_\l\rangle = -i \sum_{m\neq n}  |m_\l\rangle \langle m_\l \left| \frac{ \e_{ab}}{R} \left( \varphi_{L,0}^a \bar{\mathcal{Q}}^b + \mathcal{Q}^a \varphi_{R,0}^b\right)  - \int d\s d\tilde \s \, G(\s-\tilde \s) \J^1_t (\s) \J^2_t (\tilde \s) \right|n_\l \rangle
\ee
It is easy to argue \cite{Guica:2020eab} that the matrix elements of the first term will vanish between different eigenstates, 
and thus will drop from the sum\footnote{Note this derivation of the flow operator is significantly easier than its $J\bar T$ \cite{Guica:2020eab} and $T\bar T$ (currently not understood) counterpart, where the  main difficulty lies is finding the projection of each of the two terms on the energy eigenstates. }. It is also easy to see, e.g. using a Fourier decomposition,  that the second term will only have non-zero matrix elements if the eigenstates are different. 
Then, in the classical limit, the flow `operator' for the energy eigenstates is simply given by

\be
\X_{\mbox{\tiny{$J \bar J$}}} = i \int d\s d\ts G(\s-\ts) \J^1_t (\s) \J^2_t (\ts)
\label{flopjjbar}
\ee
%
We would  now like  to derive how  the various currents flow with respect to $\l$. For our purposes, it will be sufficient to understand this at the classical level.  If a classical current is left invariant by
%
\be
\tilde{\mathcal{D}}_\l  \equiv \p_\l   - i \{ \X_{\mbox{\tiny{$J \bar J$}}}, \; \cdot\; \} 
\ee
then we will assume this implies that at the quantum level, it will flow in the same way, \eqref{jjbflee}, as the energy eigenstates. Introducing the total momentum operator 

\be
\medpi^a = \mathcal{Q}^a +\bar{\mathcal{Q}}^a = \sqrt{k} \int d\s \,\pi^a 
\ee
%
we find that the various currents satisfy
\be
\tilde{\mathcal{D}}_\l \J_L^a = \frac{k}{2R} \e^{ab} {\medpi}_b \;, \;\;\;\;\;\; \tilde{\mathcal{D}}_\l \J_R^a = -\frac{k}{2R} \e^{ab} \medpi_b \;, \;\;\;\;\;\; \tilde{\mathcal{D}}_\l \H_L = \frac{1}{R} \e_{ab} \J_L^a \medpi^b \;, \;\;\;\;\;\; \tilde{\mathcal{D}}_\l \H_R = - \frac{1}{R} \e_{ab} \J_R^a \medpi^b \label{flcurrjjb}
\ee
Consequently, the following combinations 

\be
\tilde \H_L \equiv \H_L - \frac{ \l}{R} \e_{ab} \J_L^a \medpi^b + \frac{\l^2 k}{4 R^2} (\medpi_a)^2 \;, \;\;\;\;\;  \tilde \J_L^a \equiv \J_L^a - \frac{\l k}{2 R}  \e_{ab}\medpi^b
\ee

\be
\tilde \H_R \equiv \H_R + \frac{ \l}{R} \e_{ab} \J_R^a \medpi^b + \frac{\l^2 k}{4 R^2} (\medpi_a)^2 \;, \;\;\;\;\;  \tilde \J_R^a \equiv \J_R^a + \frac{\l k}{2 R}  \e_{ab}\medpi^b
\ee
flow in the same way as the energy eigenstates. In terms of the (dimensionless) Fourier modes of these generators, now seen as operators,  we have 


\be
\tilde J_m^a= J_m^a - \frac{ k \, \eta^a}{2}  \d_{m,0}\;, \;\;\;\;\; \tilde L_m = L_m - \eta_a J^a_m +  \frac{ k \eta_a \eta^a }{4}\d_{m,0}  \nonumber
\ee

\be
\tilde{\bar J}_m^a= \bar J_m^a - \frac{ k \, \bar \eta^a}{2}  \d_{m,0}\;, \;\;\;\;\; \tilde{\bar L}_m = \bar L_m - \bar \eta_a \bar J^a_m +  \frac{ k \eta_a \eta^a }{4}\d_{m,0} \label{fllmgenjjb}
\ee
where we introduced the operator-dependent spectral flow parameter

\be
\eta_a = \l \e_{ab} \medpi^b =-  \bar \eta_a  \label{spflopjjb}
\ee
Thus, we find that   in $J^1\wedge J^2$ - deformed CFTs, there exist two interesting bases for the symmetry generators, which are related by an operator-dependent spectral flow.  One  basis consists of the generators $L_m, J_m, \bar L_m$ and  $ \bar J_m$, which directly implement conformal and affine $U(1)$ transformations. 
The other  basis consists of the generators $\tilde L_m, \tilde J_m$  and their right-moving counterparts, which have the property  that they flow with $\l$ in the same way as the energy eigenstates, namely

\be
|n_\l \rangle = U_\l |n_{0} \rangle \;, \;\;\;\;\;\; \tilde L_m (\l) = U_\l  L_m (0) U^{-1}_\l  \label{Ufljjb}
\ee
where $U_\l = \mathcal{P} e^{ \int\mathcal{X}_{\mbox{\tiny{$J\bar J$}}}\, d\l}  \,  $ and $|n_0\rangle, L_m(0)$ are the energy eigenstates and, respectively, the symmetry generators in the undeformed CFT. This structure exactly parallels the one observed for $J\bar T$ -deformed CFTs \cite{Guica:2021pzy}. 

 In either basis, the symmetry algebra  consists of two commuting copies of the Virasoro-Kac-Moody algebra, consistently with the fact that the spectral flow parameter, even though operator-valued, commutes with all the modes of the symmetry currents.  The Hilbert space is organised into highest-weight representations of this algebra, which can be built with respect to either $L_m$ or $\tilde L_m$. Note that primary states $| h_\l\rangle$ with respect to one basis  will also be primary with respect to the other; however, the descendants in one basis will generally be a linear combination of descendants of the same level in   the other. Note also that, due to \eqref{Ufljjb}, the eigenvalues of the zero modes of the flowed generators are  independent of $\l$, and thus will equal those of the undeformed CFT

\be
\tilde L_0 | n_\l \rangle = \tilde h | n_\l \rangle\;, \;\;\;\; \tilde J_0 | n_\l \rangle = \tilde q | n_\l \rangle\;, \;\;\;\;\;\;\; \tilde{\bar L}_0 | n_\l \rangle = \tilde{\bar{ h}} | n_\l \rangle\;, \;\;\;\; \tilde{\bar J}_0 | n_\l \rangle = \tilde{\bar q} | n_\l \rangle
\ee
The flow \eqref{defdimsjjb}, \eqref{defchjjb} of the conformal dimensions and chiral charges is then explained by the relation \eqref{fllmgenjjb} between the flowed and the standard conformal generators, where the operator-dependent spectral flow parameter takes on its eigenvalue corresponding to the state under consideration.

%


\subsection{From states and  generators of symmetries to operators \label{sttoop}}

We would now like to compute correlation functions of primary operators in the $J^1 \wedge J^2$ - deformed CFT. Of course, since  the deformed theory is still a CFT and the conformal dimensions of primary operators are known \eqref{defdimsjjb}, one can immediately write down  the primary two- and three-point functions in this theory up to an overall normalization. In this section, we will show that it is in fact possible to  determine \emph{all}  the correlation functions in this model exactly in terms of the correlation functions  of the undeformed CFT. That this should have been possible is implied by the results of \cite{Cardy:2019qao} on the flow of correlation functions in $J^1 \wedge J^2$ - deformed CFTs; our method allows, in addition, to write down an entirely explicit  expression for the relation between the deformed and undeformed correlators of primary operators.

Since this exercise is supposed to serve as warm-up for the more difficult $J\bar T$ case,  we would like to phrase our computations entirely in terms of states and symmetry generators on the cylinder, which are quantities that we  have access to also in $J\bar T$ - deformed CFTs. In particular, we will do our best to avoid resorting to radial quantization or the state-operator correspondence, which have not (yet) been formulated 
for  these theories. The plan of this section is to slowly build some intuition for our construction; for the actual proposal, the reader can skip to  \eqref{defflops}.

An observable that can be straightforwardly constructed from the above building  blocks is the cylinder two-point function, seen as the overlap of an in- and an out-state created by acting with a (primary) operator on the vacuum

 \be
\O(w) | 0\rangle = e^{w h} e^{e^w L_{-1}} | h\rangle \label{ocylonvac}
\ee
In the above, $|h\rangle$ is a primary state on the cylinder, $w= \tau+i\s$ is the complex coordinate on the cylinder ($\tau=i t$) and  $L_{\pm 1} = - e^{\pm w} \p_w$, $L_0 =-\p_w$ are the  global conformal generators on the cylinder,  which  satisfy the $SL(2,\mathbb{R})$ algebra with the usual conventions. There is a completely analogous contribution from  the right-moving side, which we do not write to avoid cluttering.

The  equation above is derived in several steps: first, one uses the state-operator correspondence to  map the primary state on the cylinder to a primary operator inserted at the origin of  the plane $|h\rangle \r  \O_{pl}(0) | 0 \rangle $. This may be understood as the  definition of the primary operator.   Next, one can define an operator at an arbitrary location $z$ on the plane by translating\footnote{Note that this is not a unitarily-implemented translation, even though the prefactor   does take the familiar form 
$$
 z L_{-1} + \bar z \bar L_{-1}  = - (z \p_z + \bar z \p_{\bar z})= - (t \p_t + x \p_x) = i H t - i P x  \;\;\;\;\mbox{with} \;\;\;\; H= i \p_t\;, P= - i \p_x\;, \; z= x+it,$$
 because $L_{-1}$ is not Hermitean in radial quantization, and thus $H,P$ are not, either. 
In fact, a translation is not a symmetry of the CFT in radial quantization, because the latter singles out a special point - the origin of the plane - where operators are inserted. The fact that $\O_{pl}(z)$ takes the  form quoted  in the text  is implied by the Ward identities \eqref{wardidz} associated with translations, which are independent of the quantization we choose \cite{Simmons-Duffin:2016gjk}. }
it with $L_{-1}^{pl} = - \p_z$, i.e. $\O_{pl}(z) = e^{z L_{-1}^{pl}} \O_{pl}(0) e^{-z L_{-1}^{pl}}$. 
 When acting on the vacuum, which is annihilated by the right $L_{-1}$ factor,  we obtain the plane analogue of \eqref{ocylonvac}, $\O_{pl}(z) | 0 \rangle = e^{z L_{-1}^{pl}} |h\rangle $. The final step is to map the resulting expression back to the cylinder via $z=e^{w}$
, using the fact that  in radial quantization, $L_{-1}^{pl}$  is identified with its counterpart on the cylinder, as well as the relation  $
 \O_{(cyl)} (w) = e^{w h} \O_{pl}(z)$, which follows from the transformation
properties \eqref{primcft} of primary operators under conformal transformations. Of course, almost none of these steps would hold\footnote{Some of the complications that one encounters are: i) The map from the cylinder to the plane, assuming it can be well-defined, will be field-dependent \eqref{fdeptr}, and thus $\tau \r -\infty$ and $\tau =0$ on the cylinder will not map to a fixed location and, respectively, a fixed circle on the plane. Relatedly, dilatations correspond to a field-dependent symmetry in $J\bar T$. This makes it difficult to  formulate a state-operator correspondence precisely, even if intuitively such a map may exist; ii) The generator of right-moving translations on the plane - which are standard symmetries - does not appear to be identified with $\bar L_{-1}$ on the cylinder, which implements a field-dependent transformation. This can be easily established by noting that $L_{-1}$ and $\bar L_{-1}$  on the cylinder do not commute \eqref{llbcomm} (except when $R \r \infty$), whereas they obviously do on the plane; 
  iii) Instead of mapping back to the cylinder, one could simply  attempt to compute correlation functions on the plane. However, in this case it is not clear how to define  Hermitean conjugation, given  the general lack of understanding of radial quantization in this theory.  In particular, since $L_1^{pl}$ implements a field-dependent symmetry, it is not clear whether the coordinate appearing in the out bra  should rather be  a field-dependent coordinate. Using a different quantization, such as \cite{Gillioz:2016jnn},  does not appear to help, either.    } in $J\bar T$ - deformed CFTs, but the final result is a well-defined expression on the cylinder,   which we could simply use it to \emph{define} the operators that we would like to consider.
  
  It may in fact be possible to give an interpretation  to \eqref{ocylonvac} directly on the cylinder, by thinking of the primary state as being created by an operator insertion at $\tau= -\infty$, i.e.  $|h \rangle = \lim_{\tau \r -\infty} e^{- h \tau} \O(\tau)$,  and of the exponentiated  $L_{-1}$ as implementing a conformal transformation that brings the point at $-\infty$  to finite distance. Again, one needs to be careful about the fact that $L_{-1}$ is not a Hermitean operator; however, as we show in appendix \ref{sl2id}, its action on a primary state can be reproduced by the action of a  combination of the Hermitean operators $L_{1}+L_{-1}$ and $i(L_1 - L_{-1})$, with appropriately chosen coefficients.      While this picture does help  avoid the map to radial quantization on the plane when constructing the action of these operators, it does not necessarily help  justify a definition of the form \eqref{ocylonvac} for $J\bar T$ - deformed CFTs\footnote{The reason is that  in $J\bar T$ - deformed CFTs, the transformation taking the point at $-\infty$ to finite distance is  field-dependent, and thus one may  wonder whether one  should replace the field-independent label $e^w$ in the definition \eqref{ocylonvac} by a field-dependent one. 
In fact, one can easily check that an operator defined via \eqref{ocylonvac} in $J\bar T$ - deformed CFTs lacks a number of desirable properties - for example, the action of the field-independent operator $e^{\a L_0}$ does not correspond to the translation $w \r w+ \a$ in the label of the operator, as expected, except in the $R \r \infty$ limit. For more details, see appendix \ref{unflalg}. 
 }.

 To compute correlators, we will also need the expression for the out state
\be
\langle 0 | \O(w) = e^{-w h} \langle h | e^{e^{-w} L_1} \label{outocyl}
\ee
which follows from the simple fact that on the cylinder, hermitean conjugation\footnote{In radial quantization on the plane, one has instead 
$ \langle 0 | \O(z) \equiv ( e^{ z' L_{-1}} |h\rangle)^\dag  =z^{-2h} \langle h|  e^{ L_1/z}$, using $z'=1/z$, which follows from the action of hermitean conjugation on the cylinder  and the map $z=e^w$. } sends $w = i (t +\s) \r - w$.  
Taking the overlap, one obtains 

\be
\langle \O(w_1) \O(w_2) \rangle = e^{-h w_{12}} \langle h| e^{e^{-w_1} L_1} e^{e^{w_2} L_{-1}} |h\rangle = e^{-h w_{12}}  e^{-2 h \ln (1- e^{-w_{12}})} = \left(2 \sinh \frac{w_{12}}{2}\right)^{-2h} \label{calcother}
\ee
where we used the relation \eqref{relaat}  and the primary condition. This is  of course the correct result on the cylinder, where the dimension is given by \eqref{defdimsjjb}.
%
%

As advertised, a nice feature of this method  is that it recasts the computation  of the cylinder two-point function only in terms of states and symmetry generators, which are in principle also accessible  in $J\bar T$ - deformed CFTs\footnote{It is clear that the overlap of two states of the form \eqref{ocylonvac} and \eqref{outocyl} can be evaluated also in in $J\bar T$ - deformed CFTs, since algebra of the unflowed generators is known; see e.g. appendix \ref{unflalg}. If the operators in question only depend on the left-moving coordinate, then only the commutation relations of the left-moving generators, $L_{\pm 1}$, are relevant. These are simply $SL(2,\mathbb{R})$ commutation relations, and  one can proceed exactly as above to find the deformed left-moving piece of the two-point function, which has  the expected form. We are however interested in its  behaviour on the right-moving side. There, one encounters the complication that the algebra of $\bar L_{\pm 1,0}$ does not close, but instead generates the entire Kac-Moody tower. While the calculation is still in principle doable, we will see in the sequel that this proposal is in fact \emph{not} equivalent to the one we finally settle for in the case of $J\bar T$ - deformed CFTs.
%
}. On the down side,  this method is limited to two-point functions only.  Also, it is  not clear whether \eqref{ocylonvac} provides a satisfactory definition for the primary operators in $J\bar T$ - deformed CFTs, since  we were unable to motivate this particular choice for the action of the operator. 

To proceed, it is useful to perform the calculation of the two-point function  via overlaps in a slightly different way, 
which explicitly involves the flowed generators  \eqref{fllmgenjjb}. In terms of them, the two-point function \eqref{calcother} reads
%
 %
 \be
\langle \O(w_1) \O(w_2) \rangle  = e^{-h  w_{12}}  \langle h_\l |  e^{e^{-w_1} L_1} e^{e^{w_2} L_{-1}}| h_\l\rangle = e^{-h  w_{12}}  \langle h_\l | e^{e^{-w_1}(\tilde L_{1} +  \eta^a \tilde J^a_{1})} e^{e^{w_2} (\tilde L_{-1} +  \eta^a  \tilde J^a_{-1})} | h_\l\rangle
\ee 
where  $ \eta^a = \l \e^{ab} \medpi_b$ is the spectral flow operator  \eqref{spflopjjb} and we have reinstated the label $\l$ on the state, to emphasize its flow properties.  Since $\eta^a$ commutes with all the modes of the currents and, inside this correlator, it 
 is acting on the  state $|h_\l\rangle$, then we can simply replace it by its eigenvalue  $\etao^a = \l \e^{ab} n_b$ in this state. We then observe that the states and all the operators in the above expression flow with $\l$ in exactly the same way \eqref{Ufljjb}, and thus this correlator  will be identical to the corresponding one in the undeformed CFT. In particular, its $\l$ dependence is  entirely due to the explicit  $\l$ - dependence  of $\etao^a$.  Rather than evaluating this correlator in the undeformed CFT, we will prefer to work with the flowed states and generators in the deformed theory, keeping in mind that the two computations are simply related by conjugation by the unitary operator $U_\l$, defined in \eqref{Ufljjb}. 
 
 
The correlator can now be evaluated using the   following BCH-type formula 

\be
e^{e^{w} (\tilde L_{-1} + \eta \tilde J_{-1})} e^{- e^{w}\tilde  L_{-1}}= e^{\eta \sum_{n=1}^\infty \frac{1}{n} e^{n w} \tilde J_{-n} } \label{bchllt}
\ee
derived in appendix \ref{sl2id}, which also holds if we sum over several  currents. Using this,  the overlap is 
\bea
\langle \O(w_1) \O(w_2) \rangle 
& = & e^{- h w_{12}}\langle  h_\l| e^{e^{-w_1} \tilde L_1}  e^{\eta_\O \sum_{n=1}^\infty \frac{1}{n} e^{-n w_1} \tilde J_{n} }  e^{\eta_\O \sum_{n=1}^\infty \frac{1}{n} e^{n w_2} \tilde J_{-n} } e^{e^{w_2} \tilde L_{-1}} |h_\l \rangle \label{ovcal2pf} \\
&=& e^{-(h-\tilde h) w_{12} } e^{\frac{k \eta_\O^2}{2} \sum_{n=1}^\infty \frac{1}{n} e^{-n w_{12}}}\langle \tilde \O(w_1) e^{\eta_\O \sum_{n=1}^\infty \frac{1}{n} e^{n w_2} \tilde J_{-n} }   e^{\eta_\O \sum_{n=1}^\infty \frac{1}{n} e^{-n w_1} \tilde J_{n} }  \tilde \O(w_2) \rangle\nonumber
\eea
where in the first line we have used the fact that hermitean conjugation sends $w \r -w$, in addition to its usual effect on the generators, and in the second  line we used the BCH identity $e^A e^B = e^B e^A e^{[A,B]}$, valid if $[A,B] \propto I$, to commute the exponentials of the currents. We additionally modeled the action of   $e^{e^w \tilde L_{-1}} $ on the state $| h_\l \rangle $  by the action of an auxiliary operator $\tilde \O (w)$, acting on the vacuum 

\be
\tilde \O(w) |0_\l \rangle = e^{w \tilde h} e^{e^w \tilde L_{-1}} | h_\l\rangle
\ee
This relation follows from the corresponding relation in the undeformed CFT, by conjugation with $U_\l$. The operator $\tilde \O$ is simply \emph{defined} via the relation 

\be
\tilde \O(w) \equiv  U_\l \O_{CFT} (w) U^{-1}_\l \label{findefot}
\ee
and need not correspond to a physical operator in the deformed CFT. Given this definition, it follows that $\tilde \O(w)$ satisfies the same Ward identities with the flowed currents as the corresponding quantities in the undeformed CFT, namely

\be
[\tilde J_n^a, \tilde \O(w) ] = \tilde q^a \, e^{n w} \tilde \O(w) \;, \;\;\;\;\; \;\; n \geq 0 \label{twardidcjjb}
\ee
This in turn implies that, for any coefficients $\a_n$,

\be
 e^{\a_n \tilde J^a_n } \tilde \O(w) e^{-\a_n \tilde J^a_n} = e^{\a_n \tilde q^a e^{n w}} \tilde \O(w) \;, \;\;\;\;\;\;\;  e^{-\a_n \tilde J^a_{-n} } \tilde \O(w) e^{\a_n \tilde J_{-n}^a} = e^{-\a_n \tilde q^a e^{-n w}} \tilde \O(w)
\ee
where the second relation is the  hermitean conjugate of the first (using $\O^\dag_q (w) = \O_{-q} (-w)$). Using these relations to commute the $\tilde J_n$ and $\tilde J_{-n} $ factors past the adjacent operators (and noting the one on the left has charge $- \tilde q$), the final answer that we obtain for the correlator is 

\be
\langle \O(w_1) \O(w_2) \rangle =e^{w_{12} (\tilde h-h)}  e^{\frac{k \eta_\O^2}{2} \sum_{n=1}^\infty \frac{1}{n} e^{-n w_{12}} +  2\eta_\O \tilde q \sum_{n=1}^\infty \frac{1}{n} e^{- n w_{12}}}\langle \tilde \O(w_1) \tilde \O(w_2) \rangle =  \frac{e^{-h w_{12}} }{(1-e^{-w_{12}})^{2(\tilde h+  \etao \tilde q +\frac{k}{4} \etao^2) }}
\ee
where we used $ \sum_{n=1}^\infty \frac{1}{n} e^{-n w} = - \ln (1-e^{-w})$ and the fact that the $\tilde \O$ two-point function is identical to the one in the undeformed CFT, where the operator dimension was $\tilde h$. Thus, this method precisely reproduces the  shift of the operator dimensions due to the $J^1\wedge J^2$ deformation inside  the correlation function.

We are now ready to present our general construction. 
We define a set of ``operators'' $\tilde \O (w)$ as solutions to the flow equation

\be
\p_\l \tilde \O (w) = [\X_{\mbox{\tiny{$J\bar J$}}}, \tilde \O (w)] \label{defflops}
\ee
with the initial condition that $\tilde \O (w)_{\l =0}$ equal the CFT primary operators inserted at a point $w$ on the cylinder. One should think of these operators as being defined on the $t=0$ slice\footnote{For the original CFT operators, we should therefore write $\O(w) = e^{w L_0} \O(0) e^{- w L_0}$. }, despite the $w$ label (which has no physical meaning, except at $\l=0$). This flow equation can certainly be integrated to an equation of the  form \eqref{findefot}, though it will not in general produce a local operator in the deformed CFT (see appendix \ref{fbexjjb} for an explicit example). Nevertheless, the correlation functions\footnote{Even if our notation will be mostly euclidean, we will tacitly consider the analytic continuation to Wightman functions, in terms of which the flow picture makes sense. } of the flowed operators will be identical to those in the undeformed CFT, by virtue of the fact that they obey the same flow equation as the deformed states.  Such operators have been previously considered in \cite{Kruthoff:2020hsi}. One of their nice features is that they can  clearly  also be defined in $J\bar T$ - deformed CFTs.

Our task  is now to relate the correlation functions of the physical primary operators, $\O (w)$, in the deformed CFT  to those of the unphysical operators $\tilde \O(w)$, which simply equal the original CFT correlators. For this, we need to relate $\O(w)$ and $\tilde \O(w)$ in the deformed theory. 
 This is straightforward in $J^1 \wedge J^2$ - deformed CFTs, which are conformally
  invariant, and so primary operators must obey the usual conformal Ward identity\footnote{This follows from the usual relation \eqref{wardidz} on the plane, using $z=e^w$ and   $\O_{pl}(z) = e^{- w h} \O_{cyl} (w)$.}

\be
[L_n, \O(w)] = e^{n w}  [\,n h \O(w) +\p_w \O(w) ]\;, \;\;\;\;\;\; n \geq -1  \nonumber
\ee

\be
 [J_n^a, \O(w)] = q^a e^{n w} \O(w)\;, \;\;\;\; n \geq 0 \label{wardidcyl}
\ee
where, according  to \eqref{defdimsjjb}, \eqref{defchjjb}
\be
h = \tilde h + \eta^a_{\mbox{\tiny{$\O$}}} \tilde q_a + \frac{ k \, \eta_{\mbox{\tiny{$\O$}}}^2}{4}\;, \;\;\;\;\;\; q^a = \tilde q^a + \frac{k\,  \eta^a_{\mbox{\tiny{$\O$}}}}{2}\;, \;\;\;\;\;\; \eta^a_{\mbox{\tiny{$\O$}}} = \l \e^{ab} n_b
\ee
and similarly for the right-movers. 

On the other hand, the flow equation applied to the original Ward identity implies that  

\be
[\tilde L_n, \tilde \O(w)] = e^{n w} [\,n \tilde h \tilde \O (w) + \p_w \tilde \O (w)] \;, \;\;\;\;\;\; n \geq -1
\ee
together with \eqref{twardidcjjb}.
 Using the relationship \eqref{fllmgenjjb} between the flowed and the unflowed generators, it is easy to show that the relation between $\O (w,\bar w)$ and $\tilde \O(w,\bar w)$ is given by 

\be
\boxed{\O(w,\bar w) =  \e^{ A_{\mbox{\tiny{$\O$}}} w + B_{\mbox{\tiny{$\O$}}} \bar w} e^{\etao^a \sum_{n=1}^\infty \frac{1}{n} e^{n w} \tilde J^a_{-n} + \bar \eta^a_\O \sum_{n=1}^\infty \frac{1}{n}  e^{n \bar w} \tilde{\bar J}^a_{-n}} \tilde \O(w,\bar w) \, e^{-\eta_\O^a \sum_{n=1}^\infty \frac{1}{n} e^{- n w} \tilde J^a_n - \bar \eta_\O^a \sum_{n=1}^\infty \frac{1}{n} e^{-n \bar w} \tilde{\bar J}^a_n}} \label{reloot}
\ee
where 
\be
A_{\mbox{\tiny{$\O$}}}= \eta_a  q^a + \etao^a \tilde J^a_0-\frac{k}{4} (\etao^a)^2  -\etao^a \tilde q^a\;, \;\;\;\;\; B_{\mbox{\tiny{$\O$}}}= \bar \eta^a  \bar q_a +\baretao^a \tilde{\bar J}^a_0- \frac{k}{4} (\baretao^a)^2  -\baretao^a \tilde{\bar q}^a \label{aobojjb}
\ee
and we have reinstated the right-movers. The subscript on the operators $A_{\mbox{\tiny{$\O$}}},B_{\mbox{\tiny{$\O$}}}$ indicate that they depend on the charges of the particular operator under consideration. One should also be careful to distinguish the operators $\eta^a$ from their eigenvalue $\etao^a$, with  $[\eta^a, \O(w)] = \etao^a \O(w)$.
%
Of course, in the case at hand we have $\bar \eta_a = - \eta_a$, since the spectral flow \eqref{fllmgenjjb} acts in opposite ways on the left- and the right-movers.


One can easily check that $\O(w)$  satisfies the usual Hermiticity condition $\O_q^\dag (-w) = \O_{-q}(w)$
\be
\O_q^\dag (-w)  = e^{-\eta^a_\O \sum \frac{1}{n} e^{n w} J_{-n} } \tilde\O_{ q}^\dag (-w) e^{ \eta^a_\O \sum \frac{1}{n} e^{-n w} J^a_n - A_\O^\dag w} = e^{-A_{\O^\dag} w} \O_{-q} (w) e^{-A^\dag_{\O} w} =\O_{-q} (w) \label{odag}
\ee
%
%
where we used the fact that the charges of $\O^{\dag}$ are opposite from those of $\O$ and that $A_\O$ is hermitean. We have again dropped the right-movers, for simplicity.

Thus, we find a rather simple, closed-form relation  between the primary operators of interest and the auxiliary operators $\tilde \O$ that we defined through the flow equation. In appendix \ref{fbexjjb}, we present explicit expressions for both sets of operators for the case of  $J^1\wedge J^2$ - deformed free bosons, which make it clear that  $\tilde \O (w)$ are non-local, and thus do not correspond to physical operators that we would like to consider otherwise.  

In view of our previous discussion, 
note that the left prefactor in the relation  between $\O$ and $\tilde \O$ above  can be  understood by acting with both sides  of equation \eqref{reloot} on the vacuum, case in which it can be mapped to the relation \eqref{bchllt} between $L_{-1}$ and $\tilde L_{-1}$. Heuristically, if one thinks of the primary state as being created by the insertion of a primary operator at $\tau = -\infty$ on the cylinder, then the action of $\O(w)$ can be obtained by conformally mapping the point at infinity to finite distance using the standard conformal generator $L_{-1}$, whereas the action of $\tilde \O(w)$ is obtained by using instead  the flowed generator $\tilde L_{-1}$. However, this intuition does not help in understanding the right-hand factor in \eqref{reloot},  nor why is the spectral flow  operator  evaluated to $\etao$, even when not acting with  \eqref{reloot} on the vacuum, from either the left or the right. Therefore, while intuitively useful and correct in the particular computation of the two-point function above, the state overlap picture fails to identify  the general map between the two operators\footnote{This observation will be particularly relevant in $J\bar T$ - deformed CFTs, where the spectral flow operator does not commute with the modes of the current, and therefore it is important to establish whether it is   the operator or its eigenvalue that appears in the definition of $\O$.  }. In the following subsection, we use the relation \eqref{reloot} to compute arbitrary correlators in the $J^1\wedge J^2$ - deformed CFT.

\subsection{Correlation functions and a bootstrap check \label{jjbcorr}}

Given the expression \eqref{reloot} for the primary operators in the $J^1 \wedge J^2$ - deformed CFT  in terms of the auxiliary operators $\tilde \O (w)$,  whose correlation functions are known, computing correlation functions of $\O(w)$ becomes simply a matter of properly commuting the current modes through.
 
Let us start with the two-point function. Evaluating two copies of \eqref{reloot} in the vacuum, we find
\bea
\langle \O(w_1) O(w_2) \rangle  &= &  e^{-\sum_{i=1}^2( \frac{k}{4} \eta_i^2 + \eta_i^a \tilde q^a_i) w_i} \times  \\
& & \hspace{-1.3 cm} \times \;  \langle e^{(\eta_a q^a_1 + \eta_1^a \tilde J^a_0)w_1}\tilde \O(w_1)e^{-\eta_1^a \sum_{n=1}^\infty \frac{1}{n} e^{-n w_1} \tilde J^a_{n} } e^{(\eta_a q^a_2 + \eta_2^a \tilde J^a_0)w_2} e^{\eta_2^a \sum_{n=1}^\infty \frac{1}{n} e^{n w_2} \tilde J^a_{-n} } \tilde \O(w_2) \rangle \nonumber
\eea
where we have used the fact that the vacuum is annihilated from the left by $\tilde J^a_n$ with $n>0$ and from the right by $\tilde J_{-n}^a$. 
 Next, we note that in the above, the leftmost $\eta_a$ and $\tilde J_0^a$ will evaluate to zero, since they are acting on the vacuum; as for the middle ones, they are evaluated on the eigenstate created by $\tilde \O (w_2) | 0\rangle$, so they evaluate to $\eta_2^a$ and, respectively, $\tilde q^a_2$. The remaining manipulations are identical to those performed in the previous section, 
  and we obtain\footnote{Note that the action of a charge operator, e.g. $J_0$,  on an out state yields \emph{minus} its charge, since $J_0 | h, q \rangle = q | h,q \rangle \;\; \Rightarrow \langle h, -q | J_0 = q \langle h,-q|=-(-q)\langle h,-q|$.} 

\be
\langle \O(w_1) O(w_2) \rangle =e^{-\sum_{i=1}^2 ( \frac{k}{4} \eta_i^2 + \eta^a_i \tilde q^a_i) w_i + \eta_2^a (q^a_2 + \tilde q^a_2) w_2 } e^{- (\frac{k}{2} \eta_1^a \eta_2^a + \eta_1^a \tilde q^a_2 + \eta^a_2 \tilde q^a_1) \sum_{n} \frac{1}{n} e^{-n w_{12}}}\langle \tilde \O(w_1) \tilde O(w_2) \rangle
\ee
Performing the sum and using charge conservation, which sets $ -\eta_1 =  \eta_2 \equiv \etao$ and $-\tilde q_1 =  \tilde q_2 \equiv \tilde q$,  we find
\be
\langle \O(w_1) O(w_2) \rangle =\frac{e^{-(\etao \tilde q + \frac{k}{4} \etao^2) w_{12}}}{(1-e^{-w_{12}})^{\frac{k}{2} \etao^2 + 2 \etao \tilde q}} \cdot \left(e^{\frac{w_{12}}{2}} - e^{-\frac{w_{12}}{2}}\right)^{-2\tilde h}=  \left(e^{\frac{w_{12}}{2}} - e^{-\frac{w_{12}}{2}}\right)^{-2(\tilde h + \etao \tilde q + \frac{k}{4} \etao^2 )} \label{jjb2pf}
\ee
which is the correct result, including all the normalizations. A similar computation hols on the right.

We can use the same kind of  manipulations to compute  the three-point function
\bea
\langle \O_1(w_1) \O_2(w_2) \O_3(w_3) \rangle &=& e^{- \sum_{i=1}^3 (\frac{k}{4} \eta_i^2 + \eta_i^a \tilde q^a_i) w_i} \langle \tilde \O_1(w_1)  e^{-\eta^a_1 \sum \frac{1}{n} e^{- n w_1} \tilde J^a_n} e^{(\eta_a  q^a_2 + \eta^a_2 \tilde J^a_0 )w_2 + \eta^a_2 \sum \frac{1}{n} e^{n w_2}\tilde J_{-n}^a}  \times \nonumber \\
&& \hspace{-1.3 cm}\times \;  \tilde \O_2(w_2)\,  e^{-\eta_2^a \sum \frac{1}{n} e^{-n w_2} \tilde J^a_n} e^{(\eta_a  q^a_3 + \eta_3^a \tilde J^a_0 )w_3 + \eta^a_3 \sum \frac{1}{n} e^{n w_3} \tilde J^a_{-n}} \tilde \O_3(w_3) \rangle 
\eea
%
%
Inside this correlator, the operators $\eta_a, \tilde J_0^a$ have the following eigenvalues, from left to right:  $\eta^a= \eta^a_2+\eta^a_3 = -\eta^a_1$, $\tilde J^a_0 =-\tilde q^a_1$ and $\eta^a= \eta^a_3$, $\tilde J^a_0 =\tilde q^a_3$.  Upon commuting the current modes through, we find
\bea
\langle \O_1(w_1) \O_2(w_2) \O_3(w_3) \rangle &= & e^{-\frac{k\eta_1^2}{4}  w_1-\frac{k\eta_2^2}{4}  w_2 +\frac{k\eta_3^2}{4}  w_3- \eta_1^a \tilde q^a_1 w_1 - \eta_1^a  q^a_2 w_2+\eta^a_2 \tilde q^a_3 w_2+ \eta^a_3 \tilde q^a_3 w_3 } (1-e^{-w_{12}})^{ (\eta_2^a \tilde q^a_1 +\eta^a_1 \tilde q^a_2 +\frac{k}{2} \eta_1^a \eta^a_2)} \times \nonumber\\
& & \hspace{-1.3cm}\times \; (1-e^{-w_{23}})^{\eta^a_3 \tilde q^a_2 + \eta^a_2 \tilde q^a_3 + \frac{k}{2} \eta^a_2 \eta^a_3} (1-e^{-w_{13}})^{\eta^a_3 \tilde q^a_1 + \eta^a_1\tilde q^a_3 + \frac{k}{2} \eta^a_1 \eta^a_3} \langle \tilde \O_1 (w_1) \tilde \O_2 (w_2) \tilde \O_3 (w_3)\rangle \label{3pfojjb}
\eea
The expected form of this correlator is that of a primary three-point function on the cylinder, namely
\bea
\langle \O_1(w_1) \O_2(w_2) \O_3(w_3)\rangle  &\sim & \frac{e^{w_1 h_1 + w_2 h_2 + w_3 h_3 }}{(e^{w_1}-e^{w_2})^{h_1+h_2-h_3} (e^{w_2}-e^{w_3})^{h_2+h_3-h_1} (e^{w_1} - e^{w_3})^{h_1+h_3-h_2}} \nonumber \\
& & \hspace{-2.3cm}=\;  \frac{e^{-w_1 h_1 + w_2 (h_1-h_3) + w_3 h_3}}{(1-e^{-w_{12}})^{h_1+h_2-h_3} (1-e^{-w_{23}})^{h_2+h_3-h_1} (1-e^{-w_{13}})^{h_1+h_3-h_2}} \label{formof3pf}
\eea
 where the dimensions are given by \eqref{defdimsjjb}. It is easy to check that the exponents of the $(1-e^{-w_{ij}})$ factors in \eqref{3pfojjb} precisely match these
, since

\be
h_1 + h_2 - h_3 = \tilde h_1 + \tilde h_2 - \tilde h_3 - \eta_1^a \tilde q^a_2 - \eta_2^a \tilde  q^a_1 - \frac{k}{2} \eta^a_1 \eta^a_2
\ee
and cyclic permutations thereof, where we used charge conservation $\tilde q^a_3 = -\tilde q^a_1-\tilde q^a_2$, $\eta^a_3 = -\eta^a_1 -\eta^a_2$. Moreover, it turns out that all the prefactors in \eqref{3pfojjb} combine  precisely into the numerator of \eqref{formof3pf}, with the end result that the three-point function of primaries in the $J^1\wedge J^2$ - deformed CFT is given by \eqref{formof3pf} times the OPE coefficient, $ \tilde C_{123}$, in the \emph{undeformed} CFT.  

We thus find that, while the conformal dimensions and charges shift as in \eqref{defdimsjjb}, \eqref{defchjjb}, the OPE coefficients, which contain the dynamical information of  the theory,  are \emph{unchanged} by the deformation

\be 
C_{ABC} = \tilde C_{ABC}
\ee
Finally, we work out the four-point function, with the result
\be
 \langle \O_1(w_1) \O_2(w_2) \O_3(w_3) \O_4(w_4) \rangle = e^P \prod_{i<j} (1-e^{-w_{ij}})^{\eta_j^a \tilde q^a_i + \eta^a_i \tilde q^a_j + \frac{k}{2} \eta^a_i \eta^a_j}  \langle \tilde \O_1(w_1)\tilde \O_2(w_2)\tilde \O_3(w_3) \tilde \O_4(w_4) \rangle
\ee
where the prefactor comes from evaluating  the operators $A_i$ inside the correlator, and reads
\be
P = - \sum_i  (\frac{k}{4} \eta_i^2 + \eta_i^a \tilde q^a_i) w_i- (\eta^a_1 q^a_2 + \eta_2^a \tilde q^a_1) w_2 + [(\eta^a_3 + \eta^a_4) q^a_3 + \eta^a_3 (\tilde q^a_3 + \tilde q^a_4)]w_3 + (\eta^a_4 q^a_4 + \eta^a_4 \tilde q^a_4) w_4 
\ee
Using this, one can   check that the expression for the four-point function can be simplified to 
\be
 \langle \O_1(w_1) \O_2(w_2) \O_3(w_3) \O_4(w_4) \rangle =  \prod_{i<j} \left(e^{\frac{w_{ij}}{2}}-e^{-\frac{w_{ij}}{2}}\right)^{\frac{2}{k} (q_i q_j - \tilde q_i \tilde q_j)}  \langle \tilde \O_1(w_1)\tilde \O_2(w_2)\tilde \O_3(w_3) \tilde \O_4(w_4) \rangle \label{4pfjjb}
\ee
It is also easy to see,  using charge conservation, that any correlation function of the primary operators $\O(w_i)$ will be related through a factor of exactly the same form to the corresponding correlator in the undeformed CFT. 
Thus, the correlation functions of primary operators in the deformed theory can be rather trivially expressed in terms of the undeformed correlators.

Note that the expression \eqref{4pfjjb} for the four-point function is  crossing symmetric, assuming crossing symmetry of the original CFT correlator. It is interesting to rewrite this result in the language of conformal partial waves, using the fact that in a two-dimensional CFT that possesses an affine $U(1)$ symmetry, the  Virasoro-Kac-Moody conformal partial wave\footnote{Here, $z_i, h_i,q_i$ are the positions, dimensions and respectively $U(1)$ charges of the external operators, $h,q$ are the dimension and charge (constrained by conservation) of the primary on whose family we project, and $c$ is the  central charge of the CFT.} $\mathcal{W}_{h,q} (z_i, h_i,q_i,c)$ can be written as a spectral-flow-invariant Virasoro  block contribution $\mathcal{V}_{\hat h}$ times an affine $U(1)$ block, $\mathcal{U}$  

\be
\mathcal{W}_{h,q}(z_i, h_i, q_i,c) = \mathcal{U} (z_i, q_i) \, \V_{\hat h}(z_i, \hat h_i, c-1) 
\ee
This was shown in \cite{Fitzpatrick:2015zha} for the case of a neutral exchanged operator, and in \cite{bomb,
Song:2017czq} in the general case. In the above,  $\hat h_i = h_i - q_i^2/k$ are the spectral-flow invariant pieces of the conformal dimensions and  the affine $U(1)$ block is given by

\be
\mathcal{U} (z_i, q_i)  = \prod_{i <j} z_{ij}^{\frac{2 q_i q_j}{k}}
\ee
In a given channel, the four-point function can be written as an infinite sum over  conformal partial waves corresponding to the particular Virasoro-Kac-Moody representations being exchanged

\be
\langle \O_1 (z_1) \O_2(z_2) \O_3(z_3) \O_4(z_4)\rangle = \sum_{h,q} C_{12h} C_{34h} \mathcal{W}_{h,q}(z_i, h_i, q_i,c)
\ee 
Since the effect of the $J^1\wedge J^2$ deformation is to induce a charge-dependent spectral flow transformation that leaves $\hat h$ invariant, the only change in the Virasoro-Kac-Moody blocks will come from the change in $\mathcal{U}(z_i,q_i)$, which only depends on the  charges of the external operators. Thus, the change in the conformal partial waves is   

\be
\mathcal{W}_{h,q}(z_i, h_i, q_i,c)\;\;\; \stackrel{\mbox{\tiny{$J^1\!\!\!\wedge\!\! \! J^2$}}}{\longrightarrow} \;\;\; \prod_{i<j} z_{ij}^{\frac{2 q_i q_j}{k} - \frac{2\tilde q_i \tilde q_j}{k}}\; \mathcal{W}_{h,q}(z_i, h_i, q_i,c)
\ee
irrespectively of which operator is being exchanged. Mapping this result from the plane to the cylinder and using the fact  that the OPE coefficients are unchanged by the deformation, we can immediately reproduce the change \eqref{4pfjjb} in the four-point function.  This is another way to check that  crossing symmetry is satisfied.  The  bootstrap equations of the deformed CFT are thus trivially solved, given their solution in the undeformed CFT \cite{bomb}.

\section{Primary operators in $J\bar T$ - deformed CFTs \label{primjtb}}

Armed with our understanding of the $J^1\wedge J^2$ - deformed primaries  in a language that is in principle generalizable to $J\bar T$, we would now like to present our  proposal  for  defining primary operators in $J\bar T$ - deformed CFTs. To set up the stage, we start with a few general remarks about the similarities and differences between the $J^1\wedge J^2$ and $J\bar T$ cases. 

\subsection{Setup and general remarks \label{jtbsetup}}

Let us summarize our current understanding of $J\bar T$ - deformed CFTs that is relevant for this question. On the cylinder, the $J\bar T$ - deformed  energy eigenstates flow according to  

\be
\p_\l | n_\l \rangle = \X_{J\bar T}\,  | n \rangle_\l \label{niceflb}
\ee
where $\l$ now represents the $J\bar T$ flow parameter and  $\X_{J\bar T}$ is a presumably  well-defined operator that was worked out in \cite{Guica:2020eab} at the classical level, and discussed in \cite{Guica:2021pzy} at the quantum level
.  Given $\X_{J\bar T}$, one can define two commuting sets of Virasoro-Kac-Moody generators $\tilde L_n, \tilde J_n$ and $ \tilde{\bar L}_n, \tilde{\bar J}_n$ via the flow equation 

\be
\p_\l \widetilde L_n = [\X_{J\bar T}, \widetilde L_n] 
\ee
etc., which can be shown to be conserved \cite{Guica:2021pzy}. This flow equation implies that primary states in the undeformed CFT will flow to primaries with respect to the $\tilde L_n$ and that  the eigenvalues of $\tilde L_0$, etc., are   independent of $\l$ \cite{LeFloch:2019rut}, and thus equal the corresponding eigenvalues in the undeformed CFT

\be
\tilde L_0 | h_\l \rangle = \tilde h | h_\l\rangle \;, \;\;\;\;\;\; \tilde J_0 |h_\l\rangle = \tilde q | h_\l\rangle \;,\;\;\;\;\;\;\;\tilde{\bar L}_0 | h_\l \rangle = \tilde{\bar h} | h_\l\rangle \;, \;\;\;\;\;\; \tilde{\bar J}_0 |h_\l\rangle = \tilde{\bar q} | h_\l\rangle
\ee
Thus, in terms of the flowed generators, the structure of the Hilbert space looks  the same  as that of the undeformed CFT: primary  states have the same dimensions as in the seed CFT, and descendant states can be built by acting with $\widetilde L_{-n}$, etc.  on  them.

As explained in \cite{Guica:2021pzy}, the generators that  implement (pseudo)conformal and affine $U(1)$ transformations in the deformed theory are given by 

\be
L_n = \widetilde L_n + \l H_R \widetilde J_n + \frac{\l^2 k H_R^2}{4} \d_{n,0}\;, \;\;\;\;\;\; J_n = \tilde J_n + \frac{\l k}{2} H_R \, \d_{n,0} \nonumber
\ee

\be
\bar L_n = \widetilde{\bar L}_n + \l \!:\! H_R \widetilde{\bar J}_n \!: + \frac{\l^2 k H_R^2}{4} \d_{n,0}\;, \;\;\;\;\;\;\; \bar J_n = \tilde{\bar J}_n + \frac{\l  k }{2} H_R\, \d_{n,0} \label{unflrmg}
\ee
whose relation to the $\tilde L_n$ resembles a spectral flow whose parameter is proportional to the right-moving Hamiltonian. More specifically,  the left-moving generators $L_n, J_n$ implement usual conformal and affine $U(1)$ transformations, whereas the right-moving generators $\bar L_n, \bar J_n$ (with the exception of the global generators $\bar L_0, \bar J_0$)  implement  field-dependent  conformal \eqref{fdeptr} and affine $U(1)$ transformations. 
This structure is entirely analogous to the one we have uncovered in $J^1 \wedge J^2$ - deformed CFTs; note, however, that now the spectral flow ``parameter'' does not commute with the modes of the right-moving currents, which adds a layer of complication to the problem.

Given our understanding of the states and symmetry generators on the cylinder,  the question is how to define a physical primary operator, $\O$, and construct its correlation functions. Since the theory is local and conformal on the left,  $\O$ should obey the standard  primary condition with respect to the left-moving generators $L_n , J_n$. The non-trivial part of our task is to find an appropriate notion of a ``primary condition'' also on the non-local right-moving side. This problem does not have a counterpart in $J^1 \wedge J^2 $ - deformed CFTs, where the Ward identities that primary operators must  satisfy are simply determined by the fact that the deformed theory stays a CFT.

Since $H_R$ - the generator of right-moving translations - enters the relation \eqref{unflrmg}  between the two sets of generators, it  makes sense to choose a basis of operators that  diagonalizes its action, i.e. work in momentum space.  This is also instructed by 
the non-locality of the deformed CFT, and in particular the fact that the left primary  dimensions  depend on the  eigenvalue of this operator as in \eqref{momdepdimsjtb}. Thus, a first difference with the $J^1 \wedge J^2 $ - case is that we are forced to work 
in momentum space, at least as far as the right-movers are concerned. 
For uniformity reasons, we will choose this basis on both sides. 


As in the previous section, we will construct  candidate primary operators, $\O$, based on an auxiliary operator 
 $\tilde \O$ that is defined to flow with the $J\bar T$ parameter in the same way as the energy eigenstates.  As before, we will attempt to relate the vacuum correlation functions of the candidate $\O$ to those of $\tilde \O$, which are identical to the correlation functions in the undeformed CFT, as implied by the flow equation.  Note, however, that now 
 the `flowed vacuum' state $|0_\l\rangle = U_\l |0\rangle_{\l=0}$, to which the above argument applies,   is not annihilated by one of the global $SL(2,\mathbb{R})_L$ generators on the cylinder\footnote{This  problem does not appear in $J^1\wedge J^2$ - deformed CFTs, because the $\eta^a$  in  \eqref{spflopjjb} does annihilate the flowed vacuum. }.  More specifically, the flow equation implies that $\tilde L_{-1} | 0_\l\rangle =0$ which, in terms of the unflowed generators,  translates into

\be
 L_{-1} | 0_\l \rangle =  \l J_{-1} H_R|0_\l\rangle =  \frac{2}{\l k} \left(R-\sqrt{R^2 + \frac{\l^2 k c}{24} }\right) J_{-1} |0_\lambda \rangle \label{lm1onvac}
\ee
where we used the known result \cite{Chakraborty:2018vja} for the flowed finite-size energy eigenvalues (in this case, the ground state). Thus, on the cylinder, the flowed vacuum is not $SL(2,\mathbb{R})$ invariant. While we see no obvious reason that an $SL(2,\mathbb{R})$ - invariant vacuum should not exist, this state will clearly be different from $| 0_\l \rangle$. 

The reason we would like the vacuum to be annihilated by the  $SL(2,\mathbb{R})_L$ generators is that only then do we expect the primary correlation functions to take the standard form dictated by conformal symmetry. 
  It is therefore important to construct this state of \emph{a priori} greater physical interest in  $J\bar T$ - deformed CFTs on the cylinder\footnote{ By definition, this state would be annihilated by $L_{-1}$, which  would translate, as above, into an explicitly $\l$ - dependent relation  between the action of $\tilde L_{-1}$ and $\tilde J_{-1}$ on it.  This in turn  implies that 
this state will not satisfy a simple flow equation involving $\X_{J\bar T}$, though one may be able to explicitly write it in  the  basis \eqref{niceflb}, using  $\l$ - dependent coefficients. This state would also need to satisfy an appropriate constraint  with respect to the global right-moving generators, whose algebra - which is not $SL(2,\mathbb{R})$ - is listed in appendix \ref{unflalg}.}.  Rather than 
%
%
%
%
%
addressing this interesting problem -  which appears somewhat complicated -  we will simply avoid it by   taking the $R \r \infty$ limit, which forces the two candidate vacuum states to coincide.

Our approach can thus be summarized as follows: given that out best understanding of the states, symmetry generators and their flow is on the cylinder,   we will present our general construction of the candiate primary operators and their correlation functions in this setting.  However, this construction  is only expected to yield results consistent with $SL(2,\mathbb{R})$ invariance (and its right-moving analogue) in the $R \r \infty$ limit, where one is effectively working on the plane. This limit will turn out to also  resolve  a consistency problem that we will encounter along the way.



As we already mentioned, due to the non-locality of the model, we need to work in momentum space. We will thus start by reviewing  some basic results about momentum-space  Ward identities for primary fields in a CFT, before presenting our proposal for the momentum-space primary operators.

\subsection{CFT Ward identities in momentum space}

Let us start by introducing the momentum-space operators 

\be
\O(p) = \int d w\, e^{- p w } \O(w)
\ee
where $w = i(t+\s)$ corresponds  to a Lorentzian coordinate on the cylinder. There is, as always, also a right-moving contribution, with mometum $\bar p$, so  that the spatial momentum $p -\bar p$ is quantized. Since all formulae in this subsection are identical for the  right-movers, we will omit writing them explicitly. 

Momentum-space CFT correlation functions 
can be computed by either taking the Fourier transform of the corresponding position-space Wightman functions \cite{Gillioz:2019lgs,Bautista:2019qxj}, or  by solving the conformal Ward identities directly in momentum space \cite{Gillioz:2021sce}. While these studies were concerned with momentum-space correlators of the CFT on the plane, for the problem at hand we are interested in the form of the momentum-space conformal Ward identities on the cylinder.
%
%
%
%
These can be  obtained by  Fourier-transforming the position-space commutators \eqref{wardidcyl} of the $L_n, J_n$, with the result

\be
[L_n, \O(p)] = (n(h-1) + p R) \O\left(p-\frac{n}{R}\right) \;, \;\;\;\;\;\; [J_n, \O(p)] = q\, \O\left(p-\frac{n}{R}\right) \label{cylwidp}
\ee
Despite their unusual form, these can in principle be used to fix the form of the correlation functions\footnote{For example, for a two-point function, they give a constraint of the form
\bea
0 &=& \langle 0| L_1 L_{-1} \O_1(p_1) \O_2(p_2) |0 \rangle  = [(1-h_1+p_1 R)(h_1+p_1 R) + (1-h_2 + p_2 R) (h_2+p_2 R)] \O_1(p_1) \O_2(p_2) +   \\
&& \hspace{-1.1cm} +\; (1-h_1+p_1 R)(h_2-1+p_2 R) \O(p_1+1/R) \O(p_2-1/R) + (1-h_2+p_2 R)(h_1+p_1 R-1) \O(p_1-1/R) \O(p_2+1/R)  \nonumber
\eea
where two generator insertions are necessary in order  to respect momentum conservation. A solution to the above constraint is
%
%
\be
\langle \O_1(p_1) \O_2(p_2) \rangle \sim  e^{ \pm i \pi R (p_1-p_2)/2} \Gamma(h_1+p_1 R) \Gamma (h_2+p_2 R) \;, \;\;\;\mbox{where}\;\;\; p_1 + p_2 =0\label{cylvac2pf}
\ee
which is proportional to the Fourier transform of the position space two-point function.
}.

 While it would certainly be  interesting to further explore the properties of the solutions to these  Ward identities on the cylinder, here we are mostly interested in the  limit $R \r \infty$, where they should reduce to the well-known momentum-space  Ward identities on the plane. To show how this comes about,
%
we need to relate the (dimensionless) $SL(2,\mathbb{R})$ generators on the cylinder, $L_n = - R \, e^{\frac{n w}{R}} \p_w$ for $n=\pm1, 0$ to the planar ones, $L_n^{pl} = - w^{n+1} \p_w $, as $R\r \infty$. This can be simply achieved by expanding the cylinder generators at large $R$
\be
L_0 = - R \, \p_w = R L_{-1}^{pl} \;, \;\;\;\;\; L_{\pm 1} = - R \, \p_w \mp w \p_w - \frac{w^2}{2 R} \p_w + \O(1/R^2) 
\ee
The inverse relation reads
\be
L_{-1}^{pl} = \frac{1}{R} L_0 \;, \;\;\;\;\; L_0^{pl} = \frac{1}{2} (L_1-L_{-1}) + \O(1/R^2) \;, \;\;\;\;\;\; L_1^{pl} = R (L_1 + L_{-1} -2 L_0) + \O(1/R) \label{lplvslcyl}
\ee
Expanding the momentum-space commutators  \eqref{cylwidp} of the operator $\O (p)$ with the $L_n$, we find 

\be
[L_{-1}^{pl}, \O (p)] = p\,  \O \;, \;\;\;\;\;\; [L_0^{pl}, \O(p)] = (h-1) \O - p \, \p_p \O + \O(1/R^2) \nonumber
\ee

\be
 [L_1^{pl}, \O(p)] = p \, \p_p^2 \O + 2 (1-h) \p_p \O + \O(1/R^2) \label{plwid}
\ee 
exactly as expected. One can check that the solution to these Ward identities, e.g. for the two-point function, agrees with the $R \r \infty$ limit of its cylinder counterpart \eqref{cylvac2pf}. 

%

\subsection{A proposal for primary operators in $J\bar T$ - deformed CFTs}

We are now ready to construct a set of momentum-space operators in $J\bar T$ - deformed CFTs that come as close as possible to being primary, in the sense of  \eqref{cylwidp}. As we explained, we will start by working on the cylinder and then  take the $R \r \infty$ limit in which, at least for a CFT,  the momentum-space Ward identities and their solutions seamlessly translate  from the cylinder to the plane. 

We start by introducing again a set of operators $\tilde \O (w,\bar w)$ that formally satisfy the flow equation 

\be
\p_\l \tilde \O (w,\bar w) = [\X_{J\bar T}, \tilde \O (w,\bar w)]
\ee
with the initial condition that they equal the local operator $\O(w,\bar w)$ in the undeformed CFT. This definition automatically implies that the correlation functions of $\tilde \O(w,\bar w)$ will be identical to those in the undeformed CFT, irrespectively of the non-locality of the deformed theory, provided we evaluate them in  the flowed vacuum state, $|0_\l\rangle$. We re-emphasize that these operators - previously discussed in \cite{Kruthoff:2020hsi}  - are just auxiliary, formal constructs with no particular physical significance, as they do not correspond to local operators even on the local, left-moving side. In particular, $w, \bar w$ are simply  labels, corresponding  to the position of the initial local CFT operator that we flow, which have no particular meaning in the deformed theory.  

Since both $\tilde \O(w,\bar w)$ and $\tilde L_n, $ etc. flow in the same way with $\X_{J\bar T}$, it follows that $\tilde \O (w,\bar w)$ will satisfy the usual Ward identities

\be
[\tilde L_n, \tilde \O(w,\bar w)] = e^{n w} [n \tilde h \tilde \O (w,\bar w) + \p_w \tilde \O(w,\bar w)] \;, \;\;\;\;\;\; [\tilde J_n, \tilde \O(w,\bar w)] = e^{n w} \tilde q \, \tilde \O(w,\bar w) 
\ee

\be
[\tilde{\bar L}_n, \tilde \O(w,\bar w)] = e^{n \bar w} [n \tilde{\bar h} \tilde \O (w,\bar w) + \p_{\bar w} \tilde \O(w,\bar w)] \;, \;\;\;\;\;\; [\tilde J_n, \tilde \O(w,\bar w)] = e^{n \bar w} \tilde{\bar q}\, \tilde \O(w,\bar w) \label{otwid}
\ee
where, again, $w,\bar w$ are just labels inherited from the undeformed theory.

As explained at the beginning of this section, our candidate primary operators  should be constructed in momentum space, and  so they satisfy

\be
[H_R, \O(p,\bar p)] = \bar p \, \O(p,\bar p) \label{commhro}
\ee
In addition, they should  be primary with respect to the unflowed left-moving  generators $L_n, J_n$ 
\be
L_n = \tilde L_n + \eta \tilde J_n + \frac{k \eta^2}{4} \d_{n,0} \; , \;\;\;\;\; J_n = \tilde J_n + \frac{k \eta}{2} \d_{n,0} \;, \;\;\;\;\;\; \eta \equiv \l H_R
\ee
with the expected eigenvalues 
\be
h = \tilde h + \etao \tilde q + \frac{k \etao^2}{4} \;, \;\;\;\;\;\;\; q = \tilde q + \frac{k}{2} \etao\;, \;\;\;\;\;\; \etao = \l \bar p \label{flhqjtb}
\ee
This simply translates into the constraint \eqref{cylwidp}, where $h, q$ are given above.  The reason we introduced the above  notation is to highlight the similitude with the $J^1 \wedge J^2$ example. 
 
Given our experience with the $J^1 \wedge J^2$ deformation, we can easily construct a solution for the left-moving piece of $\O(p,\bar p)$ that satisfies  \eqref{cylwidp}. 
%
%
As for its right-moving piece, we expect a similar constraint to hold in terms of the right-moving generators $\bar L_n, \bar J_n$; however, given that these are not standard conformal generators,
we do not know exactly what relation to impose. We thus resort to simply guessing an appropriate right-moving factor, work out its properties, and then justify our choice \emph{a posteriori} via its rather reasonable predictions in the $R \r \infty$ limit.

Our proposed definition  of  an operator that has the required commutation relations with  the left-moving generators and possibly reasonable commutators with the right-moving ones is

\be \boxed{
\O(p,\bar p) = \int d w d \bar w \, e^{-p w- \bar p \bar w } \e^{A_\O w + B_\O \bar w} e^{ \eta_\O \sum_{n=1}^\infty \frac{1}{n} (e^{n w} \tilde J_{-n} +e^{n \bar w} \tilde {\bar J}_{-n})} \tilde \O(w,\bar w) e^{-\eta_\O \sum_{n=1}^\infty \frac{1}{n} ( e^{-n w} \tilde J_n  + e^{- n \bar w} \tilde{\bar J}_n )} }\label{otry}
\ee
where $A_{\mbox{\tiny{$\O$}}}$ and $B_{\mbox{\tiny{$\O$}}}$ are operators - which we will determine shortly - that depend on the conserved quantum numbers of $\O$ and - as in the $J^1 \wedge J^2$ case - are  assumed to be linear combinations of $\tilde J_0, \tilde{\bar J}_0$ and $H_R$. Note that, unlike in the $J^1\wedge J^2$ example, these operators do not commute  with  exponential factor that  follows.  The split between $A_{\mbox{\tiny{$\O$}}}, B_{\mbox{\tiny{$\O$}}}$ and  
the $p,\bar p$ factors we pulled   out is  of course completely arbitrary,  but will be soon fixed in a convenient fashion. For the time being, we are still working on the cylinder and have set $R=1$; the factor of the radius can be easily reinstated by dimensional analysis: $w \r w/R$ and $p \r p R$,   with $\eta$ fixed. 

Given this  explicit expression, we can simply compute the commutators of $\O(p,\bar p)$ with the various generators.  The simplest such commutator is with the left-moving current, which reads

\be
[ J_n, \O(p,\bar p)]= \left(\tilde q + \frac{k}{2} \etao\right) \O(p-n, \bar p) 
\ee
exactly as expected. Note that for $n\neq 0$, the shift in the charge comes from  the coefficient of $\tilde J_n$ in the exponent of \eqref{otry}, which  therefore needs
 to be a  number. For $n=0$, it comes from the commutator with $H_R$, which thus 
 sets  $ \etao = \l  \bar p $, with $\bar p$ defined in  \eqref{commhro}.
  
 %
  The commutator with $L_n $ reads
\bea
[L_n, \O(p,\bar p)] & = & \left[ n \left(\tilde h + \etao \tilde q + \frac{k \etao^2}{4} -1\right) + p \right] \O(p-n,\bar p) + \left(\etao \tilde J_0 +\eta  q  -A_{\mbox{\tiny{$\O$}}} - \etao \tilde q - \frac{k \etao^2}{4}\right) \O(p-n,\bar p ) + \nonumber \\
&+& (\l \bar p -\etao) \O(p,\bar p) \tilde J_n 
 \eea
 where the last term vanishes, for the reason we just stated. The primary condition with respect to $L_n$ then fixes
\be
 A_{\mbox{\tiny{$\O$}}}  = \etao \tilde J_0 + \eta q  -\etao \tilde q - \frac{k}{4} \etao^2 \label{aojtb} 
\ee 
which is entirely analogous to the  expression \eqref{aobojjb} we obtained in  $J^1 \wedge J^2$ - deformed CFTs.

We now turn to the commutation relations with the right-movers, still for $R$ finite. The commutation relations with the right-moving $U(1)$ current are

\be
[\bar J_n, \O(p,\bar p)] = \left(\tilde{\bar q} + \frac{k \etao}{2}\right) \O(p,\bar p-n) + \tilde{\bar J}_n [  \O(p,\bar p)-\O(p+a \a_n^r, \bar p + b \a_n^r)] \label{cjbo}
\ee
where $a,b$ are the coefficients  of $H_R$ inside $A_{\mbox{\tiny{$\O$}}}, B_{\mbox{\tiny{$\O$}}}$ and $\a_n $, given in \eqref{alphan},  is defined through the commutator 

\be
[\tilde{\bar J}_n, H_R] =  \tilde{\bar J}_n \a_n
\ee
The somewhat suspicious-looking operator-valued shift of the arguments in the last term is simply a shorthand for the corresponding Fourier-space expression, and follows from the contribution of terms of the form

\be
\tilde{\bar J}_n - e^{c H_R} \tilde{\bar J}_n  e^{- c H_R} = \tilde{\bar J}_n (1-e^{-c \a_n} )
\ee
with $c=a w$ or $b \bar w$ to the commutator. Interestingly, this shift affects both the left and the right-moving side. Note also  that \eqref{aojtb} implies that $a = \l q$.

Notwithstanding the second term - which will turn out to be negligible at large $R$ - \eqref{cjbo}  takes precisely the form of a CFT Ward identity between  an affine Kac-Moody current and an operator of right-moving charge
\be
\bar q = \tilde{\bar q} + \frac{k \etao}{2}
\ee 
which exactly mirrors the behaviour on the left-moving side. The fact that all the $\bar J_n$  modes have this behaviour, and not just the $n=0$ one (which corresponds to the global  right-moving charge), is an interesting output of our construction.

Let us now also fix the operator $B_{\mbox{\tiny{$\O$}}}$, e.g. by computing the commutator of $\O(p,\bar p)$ with $\tilde{\bar L}_0 = H_R(R-\l \tilde J_0 - \frac{\l^2 k}{4} H_R)$  which, using \eqref{commhro},  yields\footnote{Note we could have chosen the eigenvalue of $H_R$ in \eqref{commhro} to be different from the $\bar p$ factor appearing in the definition \eqref{otry}. This would have simply resulted in an expression for $B_{\mbox{\tiny{$\O$}}}$ that depended on both constants, as only $\bar p - B_{\mbox{\tiny{$\O$}}}$ is fixed.  }

\be
[\tilde{\bar L}_0, \O(p,\bar p)] = \left(\bar p + \l \tilde q \bar p + \frac{\l^2 k}{4} \bar p^2 - \l \tilde J_0 \bar p  - \l  q H_R \right)\O (p,\bar p)
\ee
This commutator can also be evaluated by  using \eqref{otry} and the Ward identity \eqref{otwid} for $\tilde \O$, with the result

\be
[\tilde{\bar L}_0, \O(p,\bar p)] = (\bar p -B_{\mbox{\tiny{$\O$}}} ) \O(p,\bar p)
\ee
%
Equating the two expressions, we find that the solution for $B_{\mbox{\tiny{$\O$}}}$ is identical\footnote{One can also check  that with this choice, $\O(p,\bar p)$ satisfies the expected Hermiticity conditions. } to the one  for $A_{\mbox{\tiny{$\O$}}}$, \eqref{aojtb}. This may seem a bit surprising, as we would have expected the right-moving  $B_{\mbox{\tiny{$\O$}}}$ to depend on the right-moving charges and current operators, as it does in $J^1\wedge J^2$ - deformed CFTs \eqref{aobojjb}, and not on the left-moving ones. 
%
The reason for this dependence can be traced back to  the fact that $\bar L_0$ - the dimensionless right-moving generator that is related to  $\tilde{\bar L}_0$ by spectral flow - does not equal $ R H_R$, but rather $\bar L_0 = R_v H_R $, where $R_v = R - \l (J_0-\bar J_0)$ is the field-dependent radius of the field-depedent right-moving coordinate. This field-dependent rescaling has no counterpart  in  $J^1\wedge J^2$ - deformed CFTs, and we could not see any simple modification of \eqref{otry} that would yield a $B_{\mbox{\tiny{$\O$}}}$ of the expected form, without spoiling the rather pleasing commutation relations we have obtained so far. One could, of course, consider working in terms of the field-dependent coordinate on the right-moving side - which natually involves factors of $R_v$ - but the resulting expressions  are significantly more complicated than \eqref{otry}.  

The reason that this particular expression for $B_{\mbox{\tiny{$\O$}}}$ is problematic  is due to its effect on the correlation functions of $\O(p,\bar p)$ on the cylinder, which will be discussed in the next subsection. However, as we will show, this effect is subleading at large $R$. Since, as we explained earlier,  the $R \r \infty$ limit is also needed to resolve the problem with the choice of vacuum in $J\bar T$ - deformed CFTs, we will simply continue to use the proposed expression \eqref{otry}, despite its drawbacks at finite $R$, and show that it does indeed yield very reasonable predictions as $R \r \infty$.


 
 Let us now finally compute 
the commutation relations of $\O(p,\bar p)$ with the unflowed right-moving generators, $\bar L_n$,  given in \eqref{unflrmg}. We find
\bea
[\bar L_n, \O(p,\bar p)] &=& \left[ n \left(\tilde{\bar h} + \etao \tilde{\bar q} + \frac{k \etao^2}{4} -1\right) + \bar p \right] \O(p,\bar p -n) + (\tilde {\bar L}_n + :\eta\, \tilde {\bar J}_n :) [\O(p,\bar p) - \O(p+a \a_n, \bar p + b \a_n)] \nonumber \\
&+& [(\etao-\eta) (\tilde q - \tilde {\bar q}) -\etao(\tilde J_0 - \tilde{\bar J}_0)] \O(p,\bar p-n) \label{clbo}
\eea
The first term on the right-hand side looks exactly like a momentum-space  conformal Ward identity in a CFT, where the right-moving conformal dimension of the operator is given by

\be
\bar h = \tilde{\bar h} + \etao \tilde{\bar q} + \frac{k \etao^2}{4}
\ee
Using $\etao=\l \bar p$, this  exactly corresponds to a momentum-dependent spectral flow of the right-moving dimensions. Note  that, unlike for the left-movers, this expression does \emph{not} follow from the flow of the  right-moving energy eigenvalues   on the cylinder, as the latter involve a factor of $\tilde q$, rather than $\tilde{\bar q}$.  

The second term can be written as $\bar L_n$ multiplying the term in parantheses (with $a=b=\l q$), since the latter vanishes for $n=0$. As we will argue, this term will drop out in the $R \r \infty$ limit. The last term is required by the definition of the right-moving generators, which include a factor of the field-dependent radius - e.g. for $n=0$, $\bar L_0= H_R R_v$. Since we have chosen to diagonalize $H_R$, and not $\bar L_0$, this term is a simple consequence of the non-trivial commutator of $\O$ with the winding operator appearing in $R_v$.

Let us now discuss the $R \r \infty$ limit of these commutators. We note that the ``operator-valued shift'' of $p,\bar p$ in \eqref{cjbo} and \eqref{clbo} scales with $R$ as 

\be
p + \frac{\l q \a_n}{R} \approx p + \frac{\l q n \hbar}{R^2}\;\;\;\;\; \mbox{as} \; R\r \infty
\ee
Thus, the contribution of this term to the commutators of $\O$ with $\bar L_n, \bar J_n$ scale as $1/R^2$ in the large $R$ limit. It is easy to see from \eqref{plwid} that these terms will consequently not contribute to the Ward identities on the plane, at least as far as  $\bar L_{\pm 1,0}$  are concerned. As for the last term in \eqref{clbo}, we note that it is suppressed by $1/R$ in the commutator with $\bar L_0/R \approx  L_{-1}^{pl}$, and it drops out from the combinations $\bar L_1 - \bar L_{-1}$ and $R(\bar L_1 + \bar L_{-1} - 2 \bar L_0) $, which are in principle identified with the plane generators. 

To summarize, while at finite $R$ our candidate primary operators obey precisely  CFT  Ward identities with respect to the left-moving generators and certain `CFT-like' Ward identities with respect to the right-moving ones, in the $R\r \infty$ limit all Ward identities  appear to reduce to exactly CFT ones, at least as far as the global conformal and Kac-Moody generators are concerned. This is consistent with the fact that the right-moving algebra  becomes Virasoro-Kac-Moody in the strict $R \r \infty$ limit. One should be cautious, however, about the presence of subtle contributions to the Ward identities in this  limit - related to the momentum dependence of the conformal dimensions - and thus a more careful study  is called for.

While the CFT-like form of the Ward identities \eqref{cjbo} and \eqref{clbo} looks rather appealing, especially as $R\r \infty$, note that we have by no means derived it from first principles. For that, one would  need to better understand how  operators  transform under the pseudo-conformal symmetries generated by $\bar L_n, \bar J_n$, taking into account the fact that their algebra is  not  Virasoro-Kac-Moody at finite $R$.  One can alternatively use  the Ward identities  we worked out  as a \emph{definition} of what is to be meant by a primary operator in the non-local $J\bar T$ - deformed  CFT, whose solution is, of course, \eqref{otry}. However, it seems hard to justify the choice \eqref{cjbo}, \eqref{clbo}, especially at finite $R$. 
%
%
 It is, nevertheless, intriguing that our candidate primary operators \eqref{otry}  satisfy such simple-looking Ward identities with respect to the pseudo-conformal  generators, which likely hint towards a much richer, pseudo-local structure of $J\bar T$ - deformed CFTs. This intuition is further supported by our results for the correlation functions, which are presented in the next subsection.

\subsection{Correlation functions}

The computation of correlation functions of the candidate primary operators \eqref{otry}  in $J\bar T$ - deformed CFTs proceeds in direct analogy to its $J^1\wedge J^2$ counterpart, which we detailed in section \ref{jjbcorr}. We will first evaluate the correlation functions at finite $R$, using the flowed vacuum $|0_\l \rangle$ - in which the correlators of the auxiliary $\tilde \O$ operators reduce to the ones in the undeformed CFT - and only take the $R \r \infty$ limit at the end. This way of proceeding will make it clear that the contribution associated with the non-CFT-like term in \eqref{clbo} drops out from the correlation function in the decompactification limit.

The momentum-space two-point function of the candidate primary operators reads

\bea
\langle \O_{-q}(p_1,\bar p_1) \O_q(p_2,\bar p_2) \rangle &= & \int d^2 w_1 d^2 w_2 e^{- \sum_i (p_i w_i + \bar p_i \bar w_i)} e^{-\eta_{vac} q (w_{12} + \bar w_{12}) - \etao (\tilde q - \tilde{\bar q}) \bar w_{12}}\times \nonumber \\
& & \hspace{6mm} \times \; \langle \O_{h_1,\bar h_1} (w_1,\bar w_1) \O_{h_2,\bar h_2} (w_2,\bar w_2) \rangle \label{jtb2pf}
\eea
where $  \bar p_1 = - \bar p_2= - \bar p$ by momentum conservation, imposed as usual by the  integral over the center of mass position.  The correlation function on the last line is a usual position-space CFT two-point function, with the same normalization as in the undeformed CFT, but with conformal dimensions   given by 

\be
 h_i (\bar p_i) = \tilde h_i + \l \tilde q_i \bar p_i +\frac{\l^2 k}{4} \bar p_i^2 \;, \;\;\;\;\;  \bar h_i (\bar p_i) = \tilde{\bar h}_i + \l \tilde {\bar q}_i \bar p_i + \frac{k \l^2}{4} \bar p_i^2 \label{momdephhb}
\ee
where $\tilde h_i, \tilde{\bar h}_i$ are the conformal dimensions of the respective operator in the undeformed CFT. The momentum-dependent shift in the dimensions, which corresponds to a spectral flow with parameters $\eta_i = \bar \eta_i= \l \bar p_i$ for each of the operators, occurs through exactly the same mechanism as for the  $J^1 \wedge J^2$ two-point function \eqref{jjb2pf}. The only differences with this previous case are that:   the spectral flow operator no longer annihilates the flowed vacuum, so we defined 

\be
\eta | 0_\l \rangle = \l H_R | 0_\l\rangle = \eta_{vac} | 0_\l\rangle \neq 0
\ee
where the actual eigenvalue can be read off from \eqref{lm1onvac}. This term contributes to the correlator as indicated above. The second difference is due to the expression \eqref{aojtb} for $B_{\mbox{\tiny{$\O$}}}$ which, as discussed,  depends on the left-moving $U(1)$ charges, instead of the right-moving ones. This leads to an explicit additional dependence on the winding charge, also indicated in \eqref{jtb2pf}.  

We now write the $\langle \O_1 (w_1,\bar w_1) \O_2 (w_2,\bar w_2) \rangle$ correlator in terms of its Fourier transform, $\mathcal{G}_{h_i,\bar h_i} (p,\bar p)\, \times$ $ \d^2(p_1+p_2)$, and perform the trivial integral over $w_i, \bar w_i$. We obtain
\bea
\langle \O_{-q} (-p, - \bar p) \O_q (p,\bar p) \rangle & = & \int dp' d\bar p' \d (p'-p+\eta_{vac} q) \d(\bar p' - \bar p +\eta_{vac} q + \etao (\tilde q -\tilde{\bar q})) \mathcal{G}_{h_i(\bar p), \bar h_i (\bar p)} (p',\bar p') \nonumber \\
&=& \mathcal{G}_{h_i(\bar p), \bar h_i (\bar p)} \big(p-\eta_{vac} q, \bar p - \eta_{vac} q - \etao (\tilde q -\tilde{\bar q})\big) \label{momsp2pfjtb}
\eea 
where $q$ itself depends on $\bar p$ as in \eqref{flhqjtb}.  Thus, the momentum-space 
  two-point function of the operators \eqref{otry} is precisely given (up to some trivial shifts in the arguments) by  a momentum-space CFT two-point function, but with the conformal dimensions  replaced by their momentum-dependent counterparts\footnote{
Note that when writing this expression, one should replace the dimensions by their momentum-dependent counterparts not only in the functional part of the correlator, but also in the prefactors, which  contain factors of e.g. $\Gamma(h)$.  This resonates with the behaviour of the   correlation functions discussed in the single-trace version of $T\bar T$-deformed CFTs, which are computed    using worldsheet string theory \cite{Asrat:2017tzd}.} \eqref{momdephhb}.
  Remarkably, this is exactly the same behaviour that we observed  in  \eqref{bh2pf} for scattering amplitudes  off near-extremal black holes!

%
%

Let us now study the three-point function of $\O(p,\bar p)$. We similarly obtain
\bea
\langle \O_1 (p_1,\bar p_1) \O_2 (p_2,\bar p_2) \O_3 (p_3,\bar p_3) \rangle &=& \int d^2 w_i\,  e^{- \sum_i (w_i p_i + \bar w_i \bar p_i)} e^{\eta_{vac}  \sum_i q_i (w_i + \bar w_i)} \times   \\ && \hspace{-5cm}\times \;  e^{-\eta_1 (q_1 - \bar q_1) \bar w_1 + [\eta_2 (q_3-\bar q_3) - \eta_1 (q_2-\bar q_2)] \bar w_2 + \eta_3 (q_3-\bar q_3) \bar w_3}  \langle \O_{h_1,\bar h_1} (w_1,\bar w_1) \O_{h_2,\bar h_2}(w_2,\bar w_2) \O_{h_3,\bar h_3}(w_3,\bar w_3) \rangle  \nonumber
\eea
where the three-point function appearing on the last line corresponds precisely to a CFT three-point function in position space, with conformal dimensions given by the momentum-dependent expressions \eqref{momdephhb} and is derived through exactly the same steps as for the $J^1 \wedge J^2$ - deformed three-point function \eqref{3pfojjb}. Fourier-transforming this (CFT) expression and performing the $w_i, \bar w_i$ integrals, one again obtains a result that corresponds to the original momentum-space CFT three-point function with the operator dimensions replaced by \eqref{momdephhb}, and with slightly shifted arguments, as in \eqref{momsp2pfjtb}. Note this implies that the OPE coefficients, appropriately defined, do not change with the deformation, as was the case in  $J^1 \wedge J^2$.  

The computation of higher-point functions on the cylinder proceeds in an identical manner. In the case of the four-point function, we note that while the position-space four-point function that appears in the integrand is entirely crossing symmetric, as we showed in section \ref{jjbcorr} for the analogous  $J^1 \wedge J^2$ deformation, the winding-dependent prefactors due to the unexpected form of $B_\O$ are not, and thus spoil the crossing symmetry of the result.  Consequently, our proposal is not quite correct  in finite size. 

This is however easy to fix by taking the $R \r \infty$ limit. Noting that the only way in which  these winding terms enter the correlator is through the shift of  the  argument of the momentum-space correlator, as in \eqref{momsp2pfjtb}, it is clear that they can be dropped in the $R \r \infty$ limit, since   they scale  as $\etao (\tilde q -\tilde{\bar{q}})/R$  with respect to $\bar p$. A similar comment applies to the $\eta_{vac} q $ term, which can also be dropped. The resulting four-point  functions  are simply the Fourier transform of position-space correlators of the form \eqref{4pfjjb}, which are manifestly  crossing symmetric and are entirely determined  by the corresponding four-point function in the undeformed CFT.   These correlation functions should be considered in the $R\r \infty$ limit, in which they simply become correlators on the plane. Identical comments apply to higher-point functions. 



The fact that all the correlation functions of our candidate $J\bar T$ primary operators are entirely determined by the original CFT correlators in such a strikingly simple manner strongly suggests that $J\bar T$ - deformed CFTs  possess a very similar structure  to that of standard two-dimensional CFTs, which simply awaits for the right language to be uncovered. We hope that some of the tools proposed in this article will be helpful in making progress on this interesting issue.

\section{Discussion \label{disc}}

In this article, we have argued that, despite their non-locality, $J\bar T$ - deformed CFTs do allow for a notion of primary operators with respect to the generators of the field-dependent symmetries that act  on the non-local side. 
We moreover showed how
 to compute  arbitrary correlation functions of these operators \emph{exactly} in terms of the correlators of the undeformed CFT. These correlation functions appear  consistent (i.e., they obey  crossing symmetry) in the decompactification limit. 

As discussed in the introduction, that special non-local theories may posses a structure that is sufficiently rigid to completely fix the form of low-point correlation functions is extremely interesting, as such theories could provide a microscopic dual to generic near-extremal (and, possibly, also non-extremal \cite{Baggio:2012db}) black holes.  It thus seems worthwhile  to better understand this structure, as well as its possible generalizations, and compare it to the results of scattering in black hole backgrounds.

A basic question is to understand from ``first principles'' the Ward identities that primary operators should satisfy, by relating them to  the expected (position-space) transformation properties of the operator under  field-dependent coordinate transformations. This ``first principles'' derivation  should also be able to determine whether there are corrections to the primary operator that involve the field-dependent coordinate, an effect that we could perhaps not see due to the large $R$ limit.

A related task is to directly work out the general constraints that these Ward identities impose on correlation functions, i.e. without appealing to the auxiliary construction involving the $\tilde \O$ operators, but rather paralleling the  usual argument   used for  standard CFTs. This question can be asked either on the cylinder or on the plane, and each  case presents its own challenges: on the cylinder,  one first needs to undestand the properties of the $SL(2,\mathbb{R})$ - invariant vacuum which, as explained, is different from the flowed one, and thus the result for the correlation functions could be rather different from those discussed in the previous section; on the plane, the momentum-space Ward identities should receive  contributions from the explicit momentum dependence of the conformal dimensions, which is not clear how to recover from the $R \r \infty$ limit.  Nonetheless, the final result that we have obtained for the correlation functions suggest that one should obtain as many constraints as there are in usual CFTs, though the language in which they are expressed may be different. 

On the more technical side, an interesting issue that we have encountered concerns the possible definitions of a vacuum state for $J\bar T$ - deformed CFTs on a cylinder, and its $SL(2,\mathbb{R})$ invariance properties. For the two possible choices of vacuum we discussed in section \ref{jtbsetup}, it would be interesting to work out their definition and exact relation, both at finite radius and in the $R \r \infty$ limit.
%
%
Another interesting technical question is to understand how the construction presented in this article works in the specific case of conserved currents,
 for example the stress tensor, and how to express these symmetry currents in terms of the associated conserved charges. This may also clarify how the field-dependent coordinate, which is essential for the definition of the charges, may fit in with the flow equations and momentum-space picture for the operators used herein.

Finally, it would be very interesting to extend these results to other special non-local theories, such as $T\bar T$ - deformed CFTs. An essential input for our present construction was the existence, in $J\bar T$ - deformed CFTs, of two different bases for the right-moving symmetries:  as a set of generators that flow in the same way as the energy eigenstates or, as the generators that directly  implement pseudoconformal transformations. In $T\bar T$ - deformed CFTs, only the first set have been shown to exist at the full quantum level \cite{Guica:2021pzy}; as for the  generators of field-dependent symmetries, they are currently understood only at a classical level and on the plane \cite{infconf}. One may nevertheless hope that the needed relation between the two will eventually be found (though, most likely, it will not correspond to a spectral flow, since the existence of a $U(1)$ current is not required in this case) and can be used to define an analogous set of primary operators. It is interesting to note that the most na\"{i}ve guess - based on the analogy with  $J\bar T$  - for how an appropriately defined `primary' two-point function will be changed - namely, via a momentum-dependent shift in the conformal dimensions -  matches the behaviour found in \cite{Cardy:2019qao}. 

Other interesting extensions of this work would be to the single-trace versions of the $T\bar T$  and $J\bar T$   deformation, where one should, in addition, be able to compare the proposed definition of the primary operators with the expectation from worldsheet string theory \cite{Asrat:2017tzd,Giribet:2017imm}. One may hope that, by extending these type of symmetries and their consequences to an ever larger class of theories, one would ultimately be able to conjecture a set of axioms (e.g., for the correlation functions) that all ``dipole CFTs'' - or, more generally, all ``non-local CFTs'' - should obey, and that this definition would be general enough to capture, in a holographic sense, the near-horizon dynamics of all extremal and non-extremal black holes.



%
\subsection*{Acknowledgements}

The author would like to thank Miguel Paulos, Sylvain Ribault and especially Balt van Rees for useful conversations. She is especially grateful to Alessandro Bombini and Andrea Galliani for collaboration on a related project. This research was supported in part by the ERC starting grant 679278 Emergent-BH. 

\appendix

\section{$SL(2,\mathbb{R})$ and Kac-Moody generator identities \label{sl2id}}

In this appendix, we derive a few identities that are useful in the manipulations of section \ref{sttoop}.

\subsubsection*{$ \bullet \;\; SL(2,\mathbb{R})$ }

Given the $SL(2,\mathbb{R})$ generators $L_{\pm 1,0}$, which satisfy the usual commutation relations, we would like to find the relations between the coefficients $(a,b,c)$ and $(\tilde a, \tilde b, \tilde c)$ in

\be
     e^{a L_{-1}} e^{2b L_0} e^{c L_1} = e^{\tilde a L_1} e^{2 \tilde b L_0 }e^{\tilde c L_{-1}}
\ee
which correspond to two different ways of parametrizing the same group element.  This identity can be derived by 
 using the following representation of $L_{\pm 1,0}$

\be
L_{-1} = \left(\begin{array}{cc}0 & 1 \\0&0\end{array}\right)\;, \;\;\;\;\;\;\; 
L_{0} = \frac{1}{2} \left(\begin{array}{cc}1 & 0 \\0&-1\end{array}\right)\;,\;\;\;\;\;\;
L_{1} = \left(\begin{array}{cc}0 & 0 \\-1&0\end{array}\right)
\ee
Using this, we find

\be
e^{a L_{-1}} e^{2b L_0} e^{c L_1} =\left(\begin{array}{cc} 1 & a \\0& 1 \end{array}\right)\left(\begin{array}{cc} e^b & 0\\0& e^{-b}  \end{array}\right)\left(\begin{array}{cc} 1&0 \\ -c & 1 \end{array}\right)=\left(\begin{array}{cc}e^b - ac e^{-b} & a e^{-b} \\-c e^{-b}&e^{-b}\end{array}\right) 
\ee
and

\be
 e^{\tilde a L_1} e^{2 \tilde b L_0 }e^{\tilde c L_{-1}} = \left(\begin{array}{cc}e^{\tilde b}  & \tilde c e^{\tilde b} \\-\tilde a e^{\tilde b}&e^{-\tilde b} - \tilde a \tilde c e^{\tilde b}\end{array}\right) 
\ee
This leads to the relation

\be
b = - \ln (e^{-\tilde b} - \tilde a \tilde c e^{\tilde b}) \;, \;\;\;\;\;\; a = \frac{\tilde c}{e^{-2\tilde b} - \tilde a \tilde c} \;, \;\;\;\;\;\; c = \frac{\tilde a}{e^{-2\tilde b} - \tilde a \tilde c} \label{relaat}
\ee
We also note the identity 
\be
e^{\a L_1 + 2 \b L_0 + \g L_{-1}}\equiv  e^M = \cosh (\sqrt{\det M}) I + \sinh (\sqrt{\det M}) \frac{M}{\sqrt{\det M}} 
\ee 
which we can use to show that ($z = Re\,  z + i Im \, z = |z| e^{i\phi}$)

\be
e^{z L_{-1} - \bar z L_1} = e^{e^{i\phi} \tanh |z| L_{-1}} e^{-2 L_0 \ln \cosh |z|} e^{- e^{-i\phi} \tanh |z| L_1}
\ee
Thus, if we let $e^w = e^{i\phi} \tanh |z|$ and act on $|h\rangle$, we obtain precisely \eqref{ocylonvac}, up to certain overall factors coming from the action of the middle term. Note that $|e^w| <1$ or $Re \, w <0$, which implies that this interpretation only applies to operators that are to the past of the $\tau =0$ slice, where the state is defined. 


Another possibly useful identity is 

\be
e^{\a L_0} e^{ e^w L_{-1}} = e^{ e^{w+\a} L_{-1}} e^{\a L_0}
\ee
which can be used to check that $L_0$ implements a translation of the state created by acting with $\O(w)$ on the vacuum $\O(w) |0\rangle  = e^{wh} e^{e^w L_{-1}} | h \rangle$.

\subsubsection*{$ \bullet \;\; SL(2,\mathbb{R})$ - Kac-Moody }

A more interesting identity to derive is the following

\be
e^{e^{w} (\tilde L_{-1} + \eta \tilde J_{-1})} e^{- e^{w}\tilde  L_{-1}}= e^{\eta \sum_{n=1}^\infty \frac{1}{n} e^{n w} \tilde J_{-n} } 
\ee
claimed in the same section, where $\eta$ is a constant or an operator that commutes with all other operators that appear in this expression. 

To prove this, we first use the Baker-Campbell-Hausdorff formula, which states that

\be
e^X e^Y = e^{\sum_{n=1}^\infty \frac{1}{n} (-1)^{n-1} \sum_{r_i,s_i}\!\! c(r_i,s_i) [X^{r_1} Y^{s_1} \ldots X^{r_n} Y^{s_n}]}
\ee
where $i \in \{1, \ldots n\}$, $r_i+s_i >0$, $c(r_i,s_i)$ are some numerical coefficients determined from these numbers, and the term in brackets is a nested commutator of $X$'s and $Y$'s. In our case

\be
X= e^w (\tilde L_{-1} + \eta \tilde J_{-1}) \;, \;\;\;\; Y= - e^w \tilde L_{-1}
\ee
so $[X,Y] = \eta e^{2w} \tilde J_{-2}$, and all further commutators with either $X$ or $Y$ will decrease the level of $\tilde J_{-n}$ by one and add a multiplicative factor of $e^w$, times some numerical coefficient. Consequently, the term on the right-hand side   must take the form 

\be
e^{e^{w} (\tilde L_{-1} + \eta \tilde J_{-1})} e^{- e^{w}\tilde  L_{-1}} = 
 e^{\eta \sum_{n=1}^\infty c_n e^{n w} \tilde J_{-n} } \label{conjbch}
\ee
for some numerical coefficients $c_n$ that we will now determine. This can be done by comparing the two ways of computing the $\langle \O(w_1) \O(w_2) \rangle$ overlap presented in section \ref{sttoop}.  

One way consists of simply repeating the steps in \eqref{ovcal2pf}, using \eqref{conjbch} instead
%
\bea
\langle \O(w_1) \O(w_2) \rangle & =& e^{-h w_{12}}  \langle h_\l | e^{e^{-w_1}(\tilde L_{1} + \eta_\O \tilde J_{1})} e^{e^{w_2} (\tilde L_{-1} + \eta_\O \tilde J_{-1})} | h_\l\rangle \nonumber\\
& = & e^{w_{12} (\tilde h-h)}\langle \tilde \O(w_1)  e^{\eta_\O \sum_{n=1}^\infty c_n e^{-n w_1} \tilde J_{n} }  e^{\eta_\O \sum_{n=1}^\infty c_n e^{n w_2} \tilde J_{-n} } \tilde \O(w_2) \rangle \nonumber  \\
&=& e^{w_{12} (\tilde h-h)} e^{\frac{k \eta_\O^2}{2} \sum_n n c_n^2 e^{-n w_{12}}}\langle \tilde \O(w_1) e^{\eta_\O \sum_{n=1}^\infty c_n e^{n w_2} \tilde J_{-n} }   e^{\eta_\O \sum_{n=1}^\infty c_n e^{-n w_1} \tilde J_{n} }  \tilde \O(w_2) \rangle\nonumber \\
&=&
 e^{w_{12} (\tilde h-h)}  e^{\frac{k \eta_\O^2}{2} \sum_n n c_n^2 e^{-n w_{12}} +  2\eta_\O \tilde q \sum_n c_n e^{- n w_{12}}}\langle \tilde \O(w_1) \tilde \O(w_2) \rangle 
\eea
However, the alternate computation  we performed, using just the commutation relations of the unflowed generators, yields directly \eqref{calcother}.  Comparing the two expressions, we conclude that  $c_n = \frac{1}{n}$. 

\section{$J^1 \wedge J^2$ - deformed free  bosons \label{fbexjjb}}

In this appendix, we work out in detail the case of $J^1\wedge J^2$ - deformed free bosons, which should help concretize the general analysis of section \ref{jjbwarm}~.  Parts of this analysis have  previously appeared in \cite{bulch,Bzowski:2018pcy}.

\subsection{Classical analysis}

We start with a variation on the calculation in appendix A of \cite{Bzowski:2018pcy}. Consider the action

\be
S= - \kappa  \int dx^+ dx^- \,[\p \phi_1 \bar \p \phi_1 +\p \phi_2 \bar \p \phi_2 - \l (\p\phi_1 \bar \p \phi_2 - \bar \p \phi_1 \p \phi_2)] = \int d \s dt \, \L \label{fba}
\ee
where $x^\pm = \s \pm t$ and $\p,\bar \p = \frac{1}{2}(\p_\s \pm \p_t)$.  The components of the two conserved shift currents $J^{\a,a} = - \frac{\p \L}{\p (\p_\a \phi_a)} $ are
\be
J^1_+ = \kappa \, (\p \phi_1 + \l \p \phi_2) 
 \;, \;\;\;\;\; J^1_{-} = \kappa \, (\bar \p \phi_1 -  \l \bar \p \phi_2) 
\ee

\be
 J^2_+ = \kappa \, (\p \phi_2 -  \l \p \phi_1) \;, \;\;\;\;\; J^2_{-} = \kappa \, ( \bar \p \phi_2 + \l \bar \p \phi_1 )
\ee
Note that the action satisfies a flow equation of the Smirnov-Zamolodchikov type\footnote{Our conventions are $\e^{\s t} = \e_{t\s} =1$.}, $\p_\l \L = - \e^{\a\b}J_\a^1 J^2_{\b} $, provided we choose
\be
\kappa = \frac{1}{1+\l^2}
\ee
%
Using this, we can construct chiral currents by taking linear combinations of $J^{1,2}_\a$ and the components of the topologically conserved current
$\tilde J^a$, with

\be
\tilde J^a_+ = \p \phi_a \;, \;\;\;\;\;\;\; \tilde J^a_{-} = -  \bar \p \phi_a
\ee
A basis for these currents is

\be
J^1_L = \frac{J^1+ \k \tilde J^1 - \l \k \tilde J_2}{2} = \k \p \phi_1\;, \;\;\;\;\;\; J^1_R =  \frac{J^1-\k \tilde J^1 - \l \k \tilde J_2}{2} = \k \bar \p \phi_1
\ee

\be
J^2_L = \frac{J^2+ \k \tilde J^2 + \l\k \tilde J_1}{2} = \k \p \phi_2\;, \;\;\;\;\;\; J^2_R =  \frac{J^2-\k \tilde J^2 + \l\k \tilde J_1}{2} = \k \bar \p \phi_2
\ee
We would now like to compute the Poisson brackets of the chiral currents. For this, we work out the canonical momenta 

\be
\pi_1 = \kappa \, (\dot \phi_1 +  \l \p_\s \phi_2) \;, \;\;\;\;\;\; \pi_2 = \kappa \,(\dot \phi_2 -  \l \p_\s \phi_1)
\ee
which satisfy the canonical equal-time commutation relations $\{\phi_a (\s), \pi_b (\s') \} = \d_{ab}\, \d(\s-\s')$, and express the currents in terms of them. We find that the Poisson brackets of the above chiral currents are diagonal in this basis, but are proportional to a factor of $\k$, which represents the level of the chiral algebra. It is desirable to work instead with the combinations

\be
\J^1_{L,R} = J^1_{L,R} \pm  \l J^2_{L,R} 
\;, \;\;\;\;\;\;\; \J^2_{L,R}= J^2_{L,R} \mp  \l J^1_{L,R}
\ee
which have level one. Their expression in terms of the canonical variables is

\be
\J^1_{L,R} = \frac{1}{2} (\pi_1 \pm \phi'_1 \pm \l \pi_2)\;, \;\;\;\;\;\;\; \J^2_{L,R} = \frac{1}{2} (\pi_2 \pm \phi'_2 \mp  \l \pi_1)
\ee 
The Hamiltonian density is given by

\be
\H =  \pi_a \dot{\phi}_a - \L = 
\frac{1+\l^2}{2} (\pi_1^2+\pi_2^2) + \frac{1}{2} (\phi_1'^2 + \phi_2'^2) + \l (\phi_1' \pi_2 - \phi_2' \pi_1)
\ee
in agreement with our general result \eqref{defhamjjb}. 

\subsection{Quantum analysis}

%
%
%

The shift in the chiral charges and the energies of (primary) states on the cylinder with momenta $n^a$ and windings $w_a$  were worked out in full generality in section \ref{brev}. Using the state-operator correspondence, these states correspond to (primary) vertex operators  that carry these charges, of the form

\be
\mathcal{V}(z,\bar z) =\; : e^{ i c^a_L \phi^a_L(z) + i c^a_R \phi^a_R (\bar z)} :
\ee
where $\phi^a_{L,R}$ are the left- and right-moving pieces of the above scalar fields and $c^a_{L,R}$ are 
%
 coefficients that we would like to determine. For simplicity, we will  concentrate on the left-moving piece of the vertex operator and drop the `$L$' index. The action \eqref{fba} implies that  the OPEs of the scalars is

\be
\phi_a (z) \phi_b (z') = - \frac{\d_{ab}}{2\k} \ln (z-z') 
\ee
  The currents $\p \phi_a$ are primary with dimension $1$, as can be checked by computing their OPE with the stress tensor,  $T_{zz} = \k  \sum_a (\p \phi_a)^2$.   
%
The OPE of the above vertex operator with the chiral left currents $\J^a = \k (\p \phi^a + \l \e^{ab} \p \phi_b)$ is thus 

\be
\J^a (z) \V(0) \sim \frac{c^a +  \l \e^{ab} c_b}{2 z  } \V(0)
\ee
Equating the coefficients of $z^{-1}$ with the flowed charge $q^a$ in \eqref{defchjjb}, we find 

\be
 c^a =2 \k (q^a - \l \e^{ab} q_b)
\ee
For the right-moving coefficients, the sign of $\l$ is switched and $q^a \r \bar q^a$.  The associated dimensions are given by the (primary) OPEs 

\be
T(z) \V(0) \sim  \frac{1}{4 z^2} \sum_a\frac{c_a^2}{\k} \V(0) + \frac{1}{z} \p \V(0) = \frac{\sum_a q_a^2}{z^2} \V(0) + \frac{1}{z} \p \V(0)
\ee
The left conformal dimension of this operator is $\sum_a(q^a)^2$, in perfect agreement with \eqref{defdimsjjb}. It is useful to rewrite the  exponent as

\be
c^a \phi_a = 2 q^a \k (\phi_a + \l \e_{ab} \phi^b) \equiv 2 q^a \varphi_a 
\ee
and similarly on the right,  where $\varphi^a_{L,R}$ simply correspond to the bosonisation of $ \J^a_{L,R}$. 
 The vertex operator thus takes the form

\be
\V(z,\bar z)  
= \; :e^{2 i q^a \varphi^a_L (z) + 2 i \bar q^a \varphi^a_R (\bar z)}:
\label{vertexop}
\ee
The mode expansion of $\varphi^a_L$ is  given in terms of the modes $ J^a_n$ of $\J^a$, i.e.

\be
\varphi^a_L (z)= \varphi^{a,0}_L -i  J_0^a\,  \ln z +i  \sum_{n\neq 0} \frac{\tilde J^a_n}{n} z^{-n } \label{phiexp}
\ee
and similarly on the right, where we used the fact that $J_n^a = \tilde J_n^a$ for $n \neq 0$ to render the expression more familiar. Note that the zero modes of the left and right chiral bosons are independent.
 It is  easy to see that these operators, even inserted at zero, will flow with $\l$. To completely specify them, we need to spell out the normal ordering - that is, we put all the annihilation operators to the right of the creation ones 

\be
\V(z) = :e^{2 i q_a \varphi^a_L}: = e^{2 i q_a \varphi^{0,a}_L + 2 q_a \sum_{n=1}^\infty \frac{\tilde J_{-n}^a}{n} z^{n} } e^{2 q_a J_0^a \ln z - 2 q_a \sum_{n=1}^\infty \frac{\tilde J^a_{n}}{n} z^{-n }}
\ee
The OPE of two such operators can be computed  using the BCH formula and the current commutation relations. We find 
\bea
 \V_{q_1} (z_1) \V_{q_2} (z_2) & = &  e^{4 i q_1^a q_2^b [J_0^a, \varphi_0^b] \ln z_1  - 2q_1^a q_2^a \sum_{n=1}^\infty \frac{1}{n} (z_2/z_1)^n}\!\! : \V_{q_1} (z_1) \V_{q_2} (z_2)  : 
 = e^{2 q_1^a q_2^a (\ln z_1 + \ln (1-z_2/z_1))} \cdot \nonumber \\
 &\cdot&: \V_{q_1} (z_1) \V_{q_2} (z_2)  : = 
  (z_1-z_{2})^{2 q_1 \cdot q_2}  :\V_{q_1} (z_1) \V_{q_2} (z_2)  :
\eea
where we used $[\varphi^{0,a}, J^b_0] = \frac{i}{2} \d^{ab}$.
This yields the correct OPE of vertex operators. Note that the zero mode contribution plays an essential role in rendering the correlator translationally-invariant. 
Including the right-moving piece, we also see that  these operators are mutually local, since  $q_1^a q_2^a - \bar q_1^a \bar q_2^a$ is $\l$-independent.

We would now like to  construct the free boson realisation of the flowed operator $\tilde \V(z)$ and compare it to the above operator. At $\l=0$, this operator is simply $\V_{CFT}(z) = :e^{2 i \tilde q_a \phi_a (z)}\!:$, where $\phi_a(z)$ has a decomposition of the form \eqref{phiexp} in terms of the \emph{undeformed} current modes. At finite $\l$, $\tilde \V $ is given by integrating  the flow equation \eqref{defflops}. 
 The flow operator in this theory is  \eqref{flopjjbar}, which in terms of Fourier modes reads

\be
 \X_{J\bar J} = \sum_{n \neq 0} \frac{1}{2\pi n} (J^1_n J^2_{-n} - \bar J^1_n \bar J^2_{-n} + J^1_n \bar J^2_n - \bar J^1_n  J^2_n)
\ee
As one can see from \eqref{flcurrjjb}, the action of this operator on the non-zero modes $\phi_a^{nzm} (z)$ inside $\V_{CFT}$  is to simply turn them into modes of the deformed current, $\tilde J_n$. On the other hand, since the zero mode of the current -  which equals the $\l$ - independent  $\tilde J_0^a$ -  commutes with $\X_{J\bar J}$, it will not  flow. The scalar zero  mode part  cannot be inferred from the simple classical flow equation; however, we do know it should be simply $2 i q_a \varphi^a_0$, because the flowed state carries charge $q_a$, and the charge is carried entirely by the zero mode. Given all this, a candidate $\tilde \V (z)$ operator is 

\be
\tilde \V(z)=:\! e^{2i q_a \varphi_0^a+ 2 i \tilde q_a (\varphi_{nzm}^a(z) -i \tilde J_0^a \ln z)  } \!:
\ee
This is in perfect agreement with the relation  \eqref{reloot} between the primary and the flowed operator, upon conformally transforming from $w$ to $z$.
%
The  fact that the coefficients of the zero and non-zero modes  are  different clearly indicates that $\tilde \V (z)$ is not a local operator. On the other hand, if we compute the OPE of two such operators, we find
\be
\tilde \V(z_1) \tilde \V(z_2) = \left(1-\frac{z_2}{z_1}\right)^{2 \tilde q_1 \cdot \tilde q_2}\!\!\! e^{4 i \tilde q_1^a  \ln z_1  q_2^b [\tilde J_0^a, q_2^b \varphi_{L,0}^b + \bar q_2^b \varphi_{R,0}^b]} :\!\tilde \V(z_1) \tilde \V(z_2)\! : =  (z_1-z_{2})^{\frac{1}{2} \tilde q_1 \cdot \tilde q_2} \!:\!\tilde \V(z_1) \tilde \V(z_2)\!:
\ee
where we
 used the fact that the commutator $ [\tilde J_0^\a, \varphi_{L/R,0}^b] = -\frac{i}{2}(\d^{ab} \d_L - \frac{\l}{2} \e^{ab}) $ and that $q^a+\bar q^a=n^a$. Therefore, these operators have  the same OPE as  the original CFT local operators, even though they are not themselves local.

\section{The unflowed  $J\bar T$ algebra \label{unflalg}}

The algebra of the unflowed generators in $J\bar T$ - deformed CFTs was spelled out in \cite{Guica:2021pzy}, and is rather involved. For the purposes of this article, we would like to only collect its subalgebra that contains the global $SL(2,\mathbb{R})$ generators $L_{\pm 1, 0}$ and their right-moving counterparts $\bar L_{\pm 1, 0}$. Unlike the case of standard CFTs, here the algebra generated by these elements does not close: instead, one must include  at least the entire infinite tower of left- and right-moving affine $U(1)$ generators. In this appendix, we spell out explicitly the commutation relations of this subalgebra, which may be helpful in following  the main text. Note that, following \cite{Guica:2021pzy}, we use the notation $K_n$ instead of $J_n$ for the current modes.

The commutation relations of the unflowed generators in the deformed $SL(2,\mathbb{R})$ - Kac-Moody subsector are, starting with the right-moving sector

\be
[\bar L_1, \bar L_{-1}] = 2 \hbar \bar L_0 + \l (\bar K_{-1} \bar L_1 + \bar L_{-1} \bar K_1 + \l \bar K_{-1} \bar K_1 \a_1) \a_1
\;, \;\;\;\;\;\; [\bar L_0, \bar L_{\pm 1}] = - \bar L_{\pm 1} \a_{\pm 1} R_v 
\ee
where 
\be
\a_n = \frac{2}{k \l^2} \left(\sqrt{(R-\l Q_K)^2 + \hbar k \l^2 n} - (R-\l Q_K) \right) = \frac{n \hbar}{R-\l Q_K} + \O(\hbar^2) \label{alphan}
\ee
with $Q_K = J_0 +\frac{\l k}{2} H_R$. Note the first commutator implies that the Kac-Moody tower cannot be decoupled, since it is generated by  $K_{\pm 1} $ and $L_{\pm 1}$. Then

\be
[\bar K_m, \bar K_n] = \frac{k m \hbar}{2} \d_{m+n} - \frac{\l k}{2} \bar K_n \a_n \d_{m,0} + \frac{\l k}{2} \bar K_m \a_m \d_{n,0}
\ee

\be
[\bar L_0, \bar K_n] = - \bar K_n \a_n R_v  \;, \;\;\;\;\;\;\; [\bar L_{-1}, \bar K_n] = -n \hbar \bar K_{n-1} - \l \bar K_{-1} \bar K_n \a_n + \frac{\l k}{2} \bar L_{-1} \a_{-1} \d_{n,0}
\ee

\be
[\bar L_1, \bar K_{n}] = - n \hbar \bar K_{n+1} - \l \bar K_n \a_n \bar K_1 + \frac{\l k}{2} \bar L_1 \a_1 \d_{n,0}
\ee
The commutators with the left-moving generators are

\be
[L_0,\bar L_{\pm 1}] = \bar L_{\pm 1} ( \pm \hbar -R \, \a_{\pm 1} ) \;, \;\;\;\;\; [L_1, \bar L_{\pm 1}] = - \l K_1 \bar L_{\pm 1} \a_{\pm 1} \;, \;\;\;\;\;
[L_{-1},\bar L_{\pm 1}] = - \l  K_{-1} \bar L_{\pm 1} \a_{\pm 1} \label{llbcomm}
\ee

\be[K_m, \bar L_n] = - \frac{\l k}{2} \bar L_n \a_n \d_{m,0} \;, \;\;\;\;\;\;\; [K_m, \bar K_n] = - \frac{\l k}{2} \bar K_n \a_n \d_{m,0}
\ee
 Note in particular that $L_{-1}$ and $\bar L_{-1}$ do not commute at finite $R$. The left generators all commute with $\bar L_0$ and their  algebra is just the standard $SL(2,\mathbb{R})$ - Kac-Moody one.

The commutation relations of these generators reduce to the standard ones in the  $R \r \infty$ limit, given the scaling \eqref{alphan} of the $\a_n$.  In particular, the combinations \eqref{lplvslcyl} of the right-moving generators do satisfy an $SL(2,\mathbb{R})$ algebra in this limit, though their definition is now only valid up to $\O(1/R)$. The commutators between the right- and  the left-moving generators also vanish as $R\r \infty$.

The above commutation relations make it clear that a proposal for the way that $\O(w)$ acts on the cylinder vacuum, of the form suggested in section \ref{sttoop}

\be
\O(w,\bar w) |0\rangle \stackrel{?}{=} e^{w h + \bar w \bar h} e^{e^w L_{-1}}  e^{e^{\bar w} \bar L_{-1}} |h, \bar h \rangle  
\ee
misses many of the properties that one may want it to have. First, since $L_{-1}$ and $\bar L_{-1}$ do not commute, this is not equivalent to e.g. $\exp \left( e^w L_{-1} + e^{\bar w} \bar L_{-1} \right)$, and thus the definition of the in/out operator is ambiguous. Second, since

\be
e^{\a \bar L_0 } e^{e^{\bar w} \bar L_{-1}} e^{-\a \bar L_0} = e^{ e^{\bar w+\a R_v \a_1^r}} \bar L_{-1}
\ee
we see that $\bar L_0$, despite being the unambiguous generator of right-moving translations on the cylinder, does not induce a simple shift in the label $\bar w$ of the operator proposed above; rather, the shift is field-dependent, except in the $R \r \infty$ limit.

%
%

%

\end{document}